\documentclass[11pt, draftclsnofoot, onecolumn]{IEEEtran}

\usepackage[utf8]{inputenc}
\usepackage{pgfplots}
\pgfplotsset{compat=newest}
\usepackage{subfigure}
\usepackage{amsmath}
\usepackage{amsfonts}
\usepackage{graphicx}
\usepackage{cite}

\usepackage{bm}
\newcommand{\ma}[1]{\mathbf{ #1 }}         

\newcommand{\compl}{\mathbb{C}}        

\usepackage{color}

\newcommand{\ignore}[1]{}

\ifCLASSINFOpdf

\else

\fi

\usepackage{algorithm}
\usepackage{algpseudocode} 
\usepackage{url}

\newcommand{\revOmid}[1]{{\color[rgb]{0,0,0}#1}}

\makeatletter

\begin{document}
\title{Quantization-Aided Secrecy: FD C-RAN Communications with Untrusted Radios}
		
\author{Omid Taghizadeh,~\IEEEmembership{Member,~IEEE},~Tianyu~Yang,~\IEEEmembership{Student~Member,~IEEE}, 
Hiroki~Iimori, \IEEEmembership{Student~Member,~IEEE},~Giuseppe~Abreu, \IEEEmembership{Senior Member,~IEEE}, 
            Ali~Cagatay~Cirik,~\IEEEmembership{Member,~IEEE},~Rudolf~Mathar,~\IEEEmembership{Senior~Member,~IEEE}          
			\IEEEcompsocitemizethanks{
							\IEEEcompsocthanksitem O.~Taghizadeh is with the 5G Wireless Research Group, Lenovo Deutschland GmbH and Network Information Theory Group, Technische Universita{\"a}t Berlin, 10587 Berlin (email: taghizadehmotlagh@tu-berlin.de).
							\IEEEcompsocthanksitem T.~Yang is with the Communications and Information Theory Chair, Technische Universita{\"a}t Berlin, 10587 Berlin (email: tianyu.yang@tu-berlin.de).
				\IEEEcompsocthanksitem H.~Iimori and G.~Abreu are with the Department of Computer Science and Electrical Engineering, Jacobs University Bremen, 28759 Bremen, Germany (Email:~h.iimori@ieee.org,~g.abreu@jacobs-university.de).
\IEEEcompsocthanksitem Ali Cagatay Cirik is with Ofinno Technologies, Ofinno Technologies, Herndon, VA, USA (email: acirik@ofinno.com).
\IEEEcompsocthanksitem Rudolf~Mathar is with the Institute for Theoretical Information Technology (TI), RWTH Aachen University, Aachen, Germany (email: mathar@ti.rwth-aachen.de).
}}
%
					
\maketitle
\vspace{-11mm}
\begin{abstract}
In this work, we study a full-duplex (FD) cloud radio access network (C-RAN) from the aspects of infrastructure sharing and information secrecy, where the central unit utilizes FD remote radio units (RU)s belonging to the same operator, i.e., the trusted RUs, as well as the RUs belonging to other operators or private owners, i.e., the untrusted RUs. Furthermore, the communication takes place in the presence of untrusted external receivers, i.e., eavesdropper nodes. The communicated uplink (UL) and downlink (DL) waveforms are quantized in order to comply with the limited capacity of the fronthaul links. In order to provide information secrecy, we propose a novel utilization of the quantization noise shaping in the DL, such that it is simultaneously used to comply with the limited capacity of the fronthaul links, as well as to degrade decoding capability of the individual eavesdropper and the untrusted RUs for both the UL and DL communications. 
In this regard, expressions describing the achievable secrecy rates are obtained. An optimization problem for jointly designing the DL and UL quantization and precoding strategies are then formulated, with the purpose of maximizing the overall system weighted sum secrecy rate. Due to the intractability of the formulated problem, an iterative solution is proposed, following the successive inner approximation and semi-definite relaxation frameworks, with convergence to a stationary point. Numerical evaluations indicate a promising gain of the proposed approaches for providing information secrecy against the untrusted infrastructure nodes and/or external eavesdroppers in the context of FD C-RAN communications. 
\end{abstract}

\begin{keywords}
Information privacy, quantization, infrastructure sharing, full-duplex, MIMO, C-RAN, physical layer security. 
\end{keywords} 

\IEEEpeerreviewmaketitle 

\section{Introduction}

%

In order to satisfy the ever-increasing demand for higher data rates, diverse usage scenarios, and service coverage extension requirements~\cite{shafi20175g}, network densification is considered as an inevitable paradigm, namely increasing the number of antennas and deploying smaller and smaller cells within an intended coverage area~\cite{interdonato2019ubiquitous}. From the network architecture perspective, the Cloud Radio Access Networks (C-RAN) enable joint baseband processing at a centralized entity, namely the Cloud Unit (CU), together with the distributed deployment of the remote radio transmitters each consisting of one or more antennas, namely the Radio Units (RU)~\cite{ALRWW:14, ZZZPKV:14, RVRWW:15, 8777303 , 8709756,  8362670, add_1, add_4, add_11}. In this respect, the network benefits simultaneously from the improved performance due to the coordinated/centralized processing and scheduling at the CU front, as well as the short-distance wireless link at the RU front. Moreover, C-RAN architecture enables optimized or on-demand deployment of the RUs as well as distributed ownership of the radio infrastructure. In particular, network and spectrum sharing have been introduced as effective methods to improve the efficiency and flexibility of the communication infrastructure~\cite{7514161, 6766097}. In a C-RAN where the radio interface is relegated to distant RUs, usually with limited availability and fronthaul capacity, efficient use of the available infrastructure is crucial. However, inter-operator cooperation leads to an inherent loss of information privacy, if not properly controlled. Furthermore, guaranteeing information security remains an ongoing challenge of the wireless communication systems due to the broadcast nature of the wireless channel, which is also exacerbated due to the distributed deployment of the RUs. 

The information security of wireless communication systems is currently addressed via cryptographic approaches, at the upper layers of the protocol stack~\cite{7954591}. However, these approaches are prone to attack due to the ever-increasing computational capability of the digital processors and suffer {from the} issues regarding management and distribution of secret keys~\cite{6739367, GKJPO13}. Alternatively, physical layer security (PLS) takes advantage of the physical characteristics of the communication medium in order to provide a secure data exchange between the information transmitter and the legitimate receiver. In the seminal work by Wyner \cite{6772207}, the concept of secrecy capacity is introduced for a three-node degraded wiretap channel, as the maximum information rate that can be exchanged \textit{under the condition of perfect secrecy}. It is shown that a positive secrecy capacity is achievable when the physical channel to the eavesdropper is {weaker than} the channel to the legitimate receiver. The arguments of \cite{6772207} {have since been extended} in the directions of secrecy rate region analysis for various wiretap channel models \cite{4529282,5730586,5961840}, construction of capacity-achieving channel codes \cite{6283924,6034749, 5545658}, as well as signal processing techniques for enhancing the {secrecy capacity}, see, e.g., \cite{chen2016survey} and the references therein.
%
%
%
%
%
%

In \cite{park2017fronthaul}, a PLS approach is proposed for the DL of a C-RAN system with untrusted RUs, and later extended for a multi-operator system under privacy constraints~\cite{8437192}. The idea is to utilize the DL fronthaul quantization, jointly shaped at the CU for all RUs, as an artificially generated noise in order to reduce the decoding capability at the untrusted RUs. In another line of work, a PLS approach is proposed for the uplink (UL) of a C-RAN system in \cite{Taghi_CRAN_TVT_Untrusted, add_5}, where the CU simultaneously utilizes the trusted as well as untrusted RUs for the purpose of communication. In the latter work, the proposed PLS scheme relies on the transmission of a friendly jamming signal, additionally generated and transmitted at the RU nodes, for the purpose of reducing decoding capability at the untrusted RUs as well as the external untrusted receivers. 

In this work, we extend the previous works which are exclusively considering information secrecy of UL or downlink (DL) of a C-RAN system, to a scenario where UL and DL directions are served jointly. In particular, we consider an FD C-RAN system where the UL and DL communication directions coexist at the same channel resource thanks to the FD capability at the RU nodes. Please note that an FD transceiver is capable of transmission and reception at the same time and frequency {band}, however, suffering from the {strong self-interference (SI)} {which is caused by its own transmitter}. The developed methods for self-interference cancellation (SIC) \cite{motz2021survey, zhang2015full}, have demonstrated {practical implementations} of FD transceivers in recent years and hence motivated several studies on the FD-enabled communication systems, both from the aspects of spectral efficiency improvements~e.g.,~\cite{cirik_FD_CRAN, 7567557}, as well as the improvement of PLS benefiting from FD jamming~\cite{GKJPO13, Schober_Sec8349956}. For the studied C-RAN network, the application of the FD RUs both enable a higher spectral efficiency due to the coexistence of the UL and DL at the same channel, as well as obtaining higher information secrecy at both directions by utilizing the fronthaul quantization as a friendly jamming signal against the untrusted entities. In particular, the DL fronthaul quantization, which is traditionally implemented in order to comply with the limited fronthaul capacity in the DL direction, is used to achieve the following additional goals: \textit{\textbf{Firstly}}, the DL fronthaul quantization noise is utilized as a friendly jamming signal on the DL fronthaul links, thereby improving the information secrecy against the untrusted RUs. \textit{\textbf{Secondly}}, the DL fronthaul quantization, after transmission from the FD RUs, is utilized as a friendly jamming signal for the untrusted users (eavesdroppers) thereby improving the information secrecy in the DL against the external eavesdroppers. \textit{\textbf{Third}}, the DL fronthaul quantization noise, after transmission from the RU, is utilized as a friendly jamming signal on the untrusted RUs as well as on the external eavesdroppers for the information transmitted in the UL direction, thereby enhancing the information security in the UL direction. The main contributions of this paper are as follows: 


\begin{itemize}
\item  In the first step, we formulate the achievable network secrecy capacity in the UL and DL directions as the function of the controllable network parameters. The achievable rate boundary is based on the results obtained for the compound wiretap channels~in \cite{liang2009compound}, as well as the mechanisms for jointly shaping the DL quantization noise over multiple channels and the resulting secrecy analysis in \cite{MultivariateQuantization_1, MultivariateQuantization_2, park2017fronthaul}. Please note that this is in contrast to the prior works in~\cite{cirik_FD_CRAN, 7567557} where the DL quantization is merely used to comply with the limited capacity at the fronthaul links, or the works targeting C-RAN security \cite{park2017fronthaul,8437192, Taghi_CRAN_TVT_Untrusted} where the DL or UL directions are studied separately.

\item On the basis of the obtained expressions, an optimization strategy is proposed for jointly obtaining the transmission and quantization strategies in the DL and UL directions, with the goal of maximizing the weighted sum secrecy rate (WSSR) of the network. Due to the non-convexity of the resulting mathematical problem, an iterative solution is proposed utilizing the successive inner-approximation (SIA)~\cite{marks1978technical}, together with the semi-definite-relaxation (SDR) framework~\cite{BV:04} with guaranteed convergence to a stationary point. Furthermore, an iterative rank-reduction procedure is proposed in order to recover a feasible solution from the SDR framework, reducing the significant complexity associated with the re-adjustments for the well-known randomization techniques~\cite{BV:04, law2015general}.  
\end{itemize}
Numerical results verify the gains of the proposed use-case, including the gains obtained by utilization of the FD capability at the RUs, the gains obtained by the utilization of the DL quantization for both UL and DL, as well as the performance improvement thanks to the proposed optimization strategy. In particular, it is observed that the proposed scheme for the coexistence of the UL and DL directions \textit{leads to an improved secrecy rate, thanks to the co-utilization and optimization of the quantization noise for multiple purposes explained above}.   

The remainder of this paper is organized as follows: the studied system model is defined in Section~II. The expressions for the achievable secure information rate at UL and DL directions are obtained in Section~III. An optimization algorithm is proposed in Section~IV. The numerical evaluations are presented in Section~V. This paper is concluded in Section~VI by summarizing the main findings.      

\subsection{Mathematical Notation:}
Column vectors and matrices are denoted as lower-case and {upper-case} bold letters, respectively. The trace, Hermitian transpose, and determinant of a matrix are respectively denoted by ${\text{ tr}}(\cdot), \; (\cdot)^{H}$, and $|\cdot|$, respectively. The Kronecker product is denoted by $\otimes$. $\left\lfloor \mathbf{A}_i \right \rfloor_{i\in\mathbb{F}}$ denotes a tall matrix, obtained by stacking the matrices $\mathbf{A}_i,~i\in\mathbb{F}$. Similarly, $\left< \ma{A}_i\right>_{i\in\mathbb{F}}$ constructs a block-diagonal matrix with the blocks $\ma{A}_i$. \revOmid{$\mathbb{E}\{\cdot\}$ denotes mathematical expectation}. $\{a_k\}$ denotes the set of all values of $a_k, \forall k$. The value of $\delta_{ij}$ is equal to $1$ for $i=j$, and zero otherwise. The set $\mathcal{A} \setminus \mathcal{B}$ includes all elements of $\mathcal{A}$, excluding those elements in $\mathcal{B}$. $\bot$ indicates statistical independence. 

\section{System Model} \label{sec_model}
\begin{figure}[!t]
\begin{center}
        \includegraphics[angle=0, width=0.85\columnwidth, height=0.40\columnwidth]{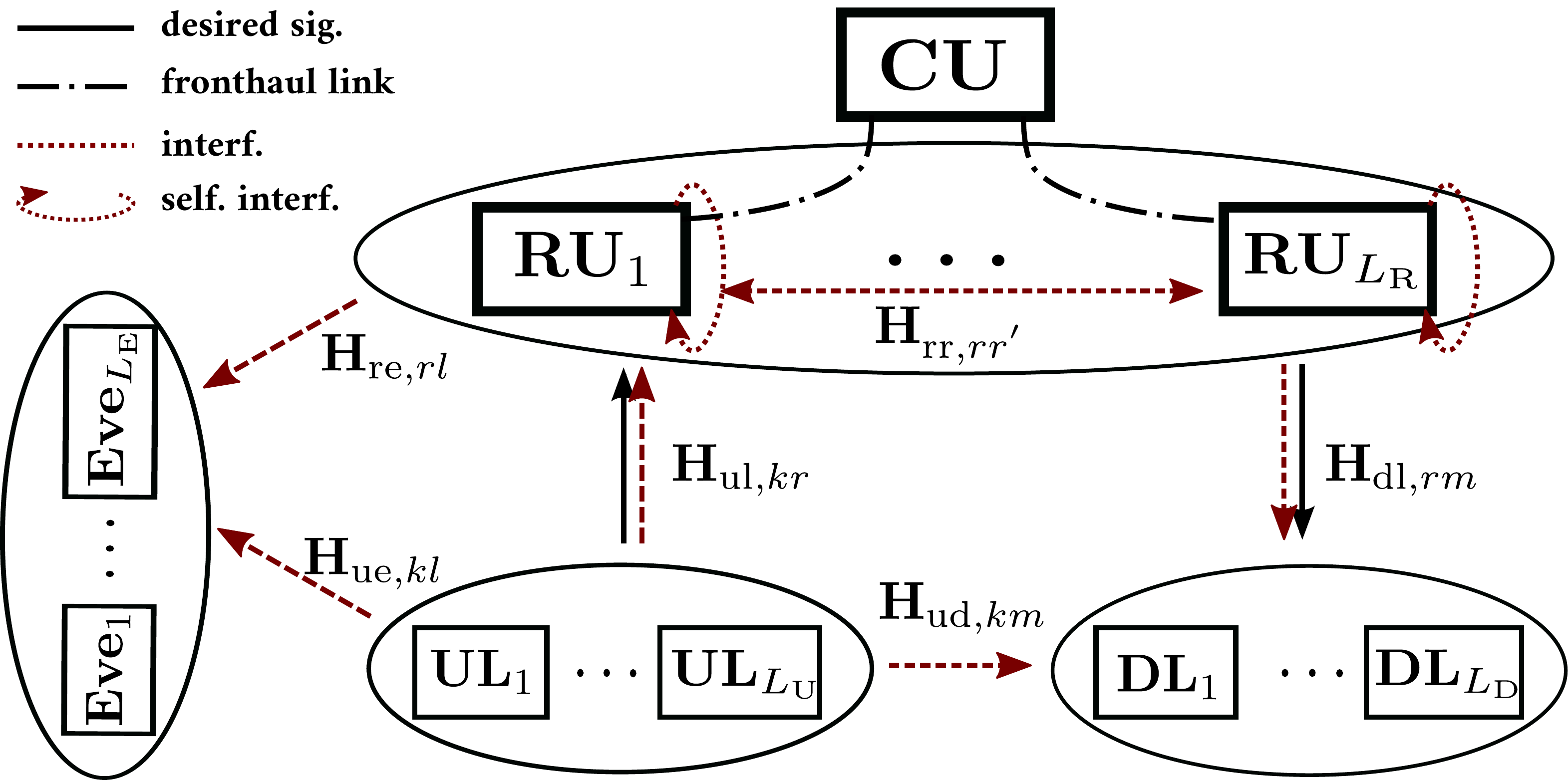}
    \caption{{The studied FD C-RAN system, including a CU unit and a group of FD-RUs simultaneously serving UL and DL users at the same frequency. Solid lines indicate the effective communication links at the UL and DL, whereas the dashed (red) arrows indicate the interference paths, see Section~\ref{sec_model} for details. }}  \label{fig_cran_sys}
    \end{center} \vspace{-0mm} 
\end{figure}

We consider an FD C-RAN communication network including a CU and a group of FD-RUs, simultaneously serving UL and DL users at the same frequency. The FD-RUs may belong to the same or a friendly operator, hence their handling of the information can be trusted, or can belong to other operators or a private owner, hence identified as an untrusted RU\footnote{Please note that as the untrusted RUs are used as part of the communication infrastructure, and hence their communication functionality can be tested and hence trusted. However, they may still store and intercept the information contained in the received waveform, hence, act as an eavesdropper.}. Furthermore, the communication is performed in the presence of the undesired information receivers, i.e., eavesdroppers. The index set of UL users, DL users, eavesdroppers, the trusted RUs and all RUs are respectively denoted as $\mathcal{U}, \mathcal{D}, \mathcal{E}, \mathcal{M}, \mathcal{R}$, such that $|\mathcal{U}|= L_{\text{U}}, |\mathcal{D}|= L_{\text{D}}, |\mathcal{E}|= L_{\text{E}},
|\mathcal{M}|= L_{\text{M}},
|\mathcal{R}|= L_{\text{R}}$. The number of transmit antennas at the RU and UL nodes is denoted as $N_r$ and $\tilde{N}_k,$ respectively, whereas the number of the receive antennas at the RUs, DL and eavesdropper nodes are denoted as ${M}_r, \tilde{M}_m$ and $\bar{M}_l,$~$\forall l \in \mathcal{E},\; m \in \mathcal{D}\;, r \in \mathcal{R},$ please see Fig.~\ref{fig_cran_sys} for a graphical description.

Each RU is connected to the CU for the UL/DL communications via a limited capacity fronthaul, where $C_{\text{ul},r}, C_{\text{dl},r}, r\in\mathcal{R},$ respectively denote the capacity of the UL and DL fronthaul links associated with the $r$-th RU. In order to comply with the limited fronthaul capacity, the UL/DL waveforms are quantized between the RUs and the CU. 

Utilizing the FD capability of the RU nodes, the UL and DL communications coexist at the same channel, which potentially improves the spectral efficiency of the system in the context of C-RAN, see \cite{ShamaiSimeone_FD_CRAN_14, cirik_FD_CRAN, 7567557, 7842356}. Furthermore, the in-band transmission and reception at the RUs enable the network to utilize the a priory-known DL quantization noise at the CU to degrade the decoding capability of the untrusted RUs, hence improving information secrecy. In this work, we employ the quasi-static block flat-fading channel model where the complex matrices $\ma{H}_{\text{ul},kr} \in \compl^{{M}_r \times \tilde{N}_{k}}, \ma{H}_{\text{dl},rm} \in \compl^{\tilde{M}_m \times {N}_{r}}, \ma{H}_{\text{ud},km} \in \compl^{\tilde{M}_m \times \tilde{N}_{k}}, \ma{H}_{\text{rr},rr^{'}} \compl^{ M_{{r}^{'}} \times N_r }, \ma{H}_{\text{ue}, kl  } \in \compl^{\bar{M}_l \times \tilde{N}_{k}}, \ma{H}_{\text{re}, rl} \in \compl^{\bar{M}_l \times N_r}$, respectively denote the user-RU, RU-user, user-user, RU-RU, UL-eavesdropper, and RU-eavesdropper channels, $\forall k \in \mathcal{U},\; l \in \mathcal{E}\;, m \in \mathcal{D},\;\forall r\neq r^{'} \in \mathcal{R}$. 

\subsubsection{Transmitted signal model}                
The CU transmit waveform for the $r$-th RU, before and after quantization is respectively denoted as                                         
\begin{align}                                        
\ma{x}_{\text{dl},r}^{\text{(CU)}} &= \sum_{m \in \mathcal{D}} \ma{W}_{m,r} \ma{s}_{\text{dl},m}, \;\; \forall r \in \mathcal{R}, \label{eq_tx_cu_dl}\\
\ma{x}_{\text{dl},r} &= \ma{x}_{\text{dl},r}^{\text{(CU)}} + \ma{q}_{\text{dl},r}, \;\; I \left(\ma{x}_{\text{dl},r}^{\text{(CU)}}; \ma{x}_{\text{dl},r} \right) \leq C_{\text{dl},r}, \;\; \forall r \in \mathcal{R},                       
\end{align}                                          
where $\ma{q}_{\text{dl},r} \in \compl^{N_r}$ is the DL quantization noise, $\ma{s}_{\text{dl},m} \sim \mathcal{CN}\left( 0 , \ma{I}_{d_{m}} \right)$ denotes the DL data symbol with dimension $d_{m}$, and $\ma{W}_{m,r} \in \compl^{N_r \times d_m}$ is the associated DL transmit precoder for the $r$-th RU. Please note that the constraint on the mutual information between the actual and the quantized waveform is necessary to comply with the limited fronthaul capacity $C_{\text{dl},r}$, see \cite{MultivariateQuantization_1, Taghi_CRAN_TVT_Untrusted}. At the UL side, the transmitted signal from each UL user is written as 
\begin{align}
\ma{x}_{\text{ul},k} &= \ma{F}_{k} \ma{s}_{\text{ul}, k}, \;\; \forall k \in \mathcal{U},
\end{align}                                        
where $\ma{F}_{k} \in \compl^{  \tilde{N}_{k} \times \tilde{d}_k}$ and $\ma{s}_{\text{ul}, k} \in \sim \mathcal{CN} \left(0, \ma{I}_{\tilde{d}_k} \right)$ are the UL transmit precoder and the vector of UL data symbols with dimension $\tilde{d}_k$, respectively. In order to comply with the limited power budget as well as the limited battery output range, the transmit power constraints at the UL users and RUs are respectively expressed as
\begin{align}                                      
\mathbb{E} \left\{ \|\ma{x}_{\text{ul},k}\|_2^2 \right\} & \leq P_{\text{ul},k}, \;\; \forall k \in \mathcal{D}, \\                                                             
\mathbb{E} \left\{ \|\ma{x}_{\text{dl},r}\|_2^2 \right\} & \leq P_{\text{dl},r}, \;\; \forall r \in \mathcal{R},
\end{align}  
where $P_{\text{dl},r}, P_{\text{ul},k}$ respectively represent maximum transmit power at the $r$-th RU and at the $k$-th UL users. 
                       
\subsubsection{Received signal model}
Consequently, the received signal at the DL users and at the RU nodes are respectively written as  
                       
\begin{align}           
\ma{y}_{\text{dl} ,m} &=  \sum_{r \in \mathcal{R}} \ma{H}_{\text{dl},rm} \ma{x}_{\text{dl},r}  +  \sum_{k \in \mathcal{U}} \ma{H}_{\text{ud},km}   \ma{x}_{\text{ul},k}  + \ma{n}_{\text{dl},m}, \;\; \forall m \in \mathcal{D}, \\
\ma{y}_{\text{ul},r} &= \sum_{k \in \mathcal{U}} \ma{H}_{\text{ul}, kr} \ma{x}_{\text{ul},k} + \sum_{r^{'} \in \mathcal{R} \setminus r} \ma{H}_{\text{rr}, {r^{'}}r} \ma{x}_{ \text{dl}, {r^{'}} }  + \ma{n}_{\text{ul},r} + \boldsymbol{\nu}_r , \;\; \forall r \in \mathcal{R},
\end{align}      
where $\ma{n}_{\text{dl},m} \sim \mathcal{CN} \left(0, N_{\text{dl},m} \ma{I}_{\tilde{M}_m}\right)$ and $\ma{n}_{\text{ul},r} \sim \mathcal{CN} \left(0, N_{\text{ul},r}\ma{I}_{M_r}\right)$ respectively denote the thermal noise at the DL user and the RU node and
$\boldsymbol{\nu}_r \in \compl^{M_r}$ represents the residual self-interference at the $r$-th RU, remaining from the self-interference cancellation at the FD RU node, please see Subsection~\ref{subsec_SRI} for more details on the self-interference cancellation methods and modeling of the residual impairments. Similarly, the received signal at the eavesdropper nodes are expressed as
\begin{align}   
\ma{y}_{\text{e},l} &= \sum_{k \in \mathcal{U}} \ma{H}_{\text{ue}, kl} \ma{x}_{\text{ul},k} + \sum_{r \in \mathcal{R}} \ma{H}_{\text{re}, rl} \ma{x}_{\text{dl}, r}  + \ma{n}_{\text{e},l}, \;\; \forall l \in \mathcal{E},  \label{eq_rx_eve}                                                   
\end{align}      
where $\ma{n}_{\text{e},l} \sim \mathcal{CN} \left(0, N_{\text{e},l} \ma{I}_{\bar{M}_l}\right)$ is the additive thermal noise at the $l$-th eavesdropper.                                          
                                                                                    
Similar to the DL waveform, in order to comply with the limited capacity of the UL fronthaul link, the quantized version of the received UL waveform is delivered to the CU, i.e.,                    
\begin{align}                                                                      
\ma{y}_{\text{ul},r}^{\text{(CU)}} &= \ma{y}_{\text{ul},r} + \ma{q}_{\text{ul},r}, \;\; I \left(\ma{y}_{\text{ul},r}^{\text{(CU)}}; \ma{y}_{\text{ul},r} \right) \leq C_{\text{ul},r}, \;\; \forall r \in \mathcal{R}, 
\end{align}                                                                       
where $\ma{q}_{\text{ul},r} \in \compl^{M_r}$ is the UL quantization noise, and the right hand-side constraint ensures that UL waveform complies with the finite fronthaul capacity in the RU-CU link.  
                
\subsubsection{Notation simplification}
For notational convenience, we define the bold-faced representation of the vector/matrix $\ma{X}$ to be the vector/matrix obtained by stacking the blocks $\ma{X}_r$ over all RUs and dropping the associated index $r$. In particular, we have $\ma{X} := \lfloor \ma{X}_r \rfloor_{r \in {\mathcal{R}}}$, such that 
\begin{equation}
    \ma{X}_r \in \{\ma{W}_{m,r},  \ma{q}_{\text{dl},r}, \ma{q}_{\text{ul},r}, \ma{x}_{\text{dl},r}, \ma{x}_{\text{ul},r}, \ma{x}_{\text{dl},r}^{\text{(CU)}}, \ma{x}_{\text{ul},r}^{\text{(CU)}}, \ma{n}_{\text{ul},r}, \boldsymbol{\nu}_r, \ma{H}_{\text{dl},rm} , \ma{H}_{\text{ul}, kr}, \ma{H}_{\text{re},rl} \}.
\end{equation}
Furthermore, the selection matrices 
\begin{align}   \label{eq_sel_matrix} 
\ma{S}_{\text{ul},r} = \left[ \ma{0}_{M_r \times \sum_{i=1}^{r-1} M_r}, \; \ma{I}_{M_r}, \; \ma{0}_{M_r \times \sum_{i = r+1}^{L_{\text{R}}} M_r} \right], \;\; \ma{S}_{\text{dl},r} = \left[ \ma{0}_{N_r \times \sum_{i=1}^{r-1} N_r}, \; \ma{I}_{N_r}, \; \ma{0}_{N_r \times \sum_{i = r+1}^{L_{\text{R}}} N_r} \right],
 \end{align} 
 are respectively used to extract the received and transmit signal associated with the $r$-th RU from the stacked array. The expressions in (\ref{eq_tx_cu_dl})-(\ref{eq_rx_eve}) can be hence reformulated as 
\begin{align}
\ma{x}_{\text{dl}}^{\text{(CU)}} &= \sum_{m \in \mathcal{D}} \ma{W}_{m} \ma{s}_{\text{dl},m}, \;\; \ma{x}_{\text{dl}}= \ma{x}_{\text{dl}}^{\text{(CU)}} + \ma{q}_{\text{dl}}, \\
\ma{y}_{\text{ul}}^{\text{(CU)}} &= \ma{y}_{\text{ul}} + \ma{q}_{\text{ul}}, \;\;
\ma{y}_{\text{ul}} = \sum_{k \in \mathcal{U}} \ma{H}_{\text{ul}, k} \ma{x}_{\text{ul},k} + \ma{H}_{\text{rr}} \ma{x}_{\text{dl}} + \ma{n}_{\text{ul}} + \boldsymbol{\nu}, \\
\ma{y}_{\text{e},l} &= \sum_{k \in \mathcal{U}} \ma{H}_{\text{ue}, kl} \ma{x}_{\text{ul},k} + \ma{H}_{\text{re},l} \ma{x}_{\text{dl}}  + \ma{n}_{\text{e},l}, \;\; \forall l \in \mathcal{E},  \label{eq_rx_eve_stacked}             
\end{align}  
where $\ma{H}_{\text{rr}} := \lfloor \ma{H}_{\text{rr}, r} \rfloor_{r \in \mathcal{R}}$ such that $\ma{H}_{\text{rr}, r}^T = \lfloor  (1-\delta_{r r^{'}}) \ma{H}_{\text{rr}, {r^{'}}r} ^T \rfloor_{r^{'} \in \mathcal{R}}$ represents the inter-RU interference channel excluding the self-interference, and $\ma{q}_{\text{ul}} \sim \mathcal{CN}\left( \ma{0}, \ma{Q}_{\text{ul}} \right)$ and $\ma{q}_{\text{dl}} \sim \mathcal{CN} \left( \ma{0}, \ma{Q}_{\text{dl}} \right)$ and $\boldsymbol{\nu}$ are respectively the vectorized quantization noise for the UL, DL, and the residual self-interference, such that $\ma{Q}_{\text{ul}} = \left< \ma{Q}_{\text{ul},r}\right>_{r\in\mathbb{R}}$. 

\subsection{Residual Self-interference} \label{subsec_SRI}
Employing the developed SIC methods in various signal domains, an FD transceiver is capable of estimating and effectively suppressing the received self-interference signal, e.g.,~\cite{Bharadia:14,YLMAC:11,BMK:13,khandani2013two}. Nevertheless, the accuracy of the employed SIC methods is limited due to the limited dynamic range at the transmit (Tx) and receive (Rx) chains, as well as the strength of the self-interference channel. To this end, it is widely known that the consideration of the limited hardware and SIC accuracy is essential in the design and performance evaluation of the FD-enabled networks~\cite{8234646, DMBS:12, DMBSR:12}. The impact of the limited Tx/Rx chain accuracy in the context of the FD transceiver has been studied in \cite{DMBS:12, DMBSR:12}, based on the prior experimentation~\cite{MITTX:98, MITRX:05, MITTX:08}, and widely used in the context of FD system design and performance analysis, e.g., \cite{8234646, DMBSR:12, XaZXMaXu:15, ALRWW:14, ZZZPKV:14, RVRWW:15, 8777303 , 8709756,  8362670}. 
In particular, the proposed model in \cite{DMBS:12} is based on the following three observations. Firstly, the collective distortion signal in each transmit/receive chain can be approximated as an additive zero-mean Gaussian term. Secondly, the variance of the distortion signal is proportional to the power of the intended transmit/received signal. And third, the distortion signal is statistically independent of the intended transmit/receive signal at each chain, and among different chains, see \cite[Subsections~B-C]{DMBS:12}. Consequently, in the studied C-RAN network, the statistics of the residual self-interference can be expressed as 
\begin{align}       
&\boldsymbol{\nu} \sim \mathcal{CN} \left( \ma{0}, \ma{\Lambda} \left( \{\tilde{\ma{W}}_m \}, \ma{Q}_{\text{dl}} \right) \right), \\
& \ma{\Lambda} \left( \{\tilde{\ma{W}}_m \}, \ma{Q}_{\text{dl}} \right) :=  \kappa \tilde{\ma{H}}_{\text{rr}} \text{diag}\left( \ma{Q}_{\text{dl}} + \sum_{m \in \mathcal{D}} \ma{W}_m \right)  \tilde{\ma{H}}_{\text{rr}}^H + \beta \text{diag} \left( \tilde{\ma{H}}_{\text{rr}} \left( \ma{Q}_{\text{dl}} + \sum_{m \in \mathcal{D}} \ma{W}_m \right)  \tilde{\ma{H}}_{\text{rr}}^H \right),
\end{align}
where $\ma{\Lambda}$ is the covariance of the residual self-interference and $\tilde{\ma{W}}_m := {\ma{W}}_m {\ma{W}}_m^H$ is the DL transmit covariance associated with the $m$-th user. In the above expressions, $0<\kappa,\beta \ll 1$ are respectively the transmit and receive distortion coefficients, relating the transmit signal power to the residual self-interference variance and $\tilde{\ma{H}}_{\text{rr}} = \lfloor \lfloor  \ma{H}_{\text{rr}, {r^{'}}r} ^T \rfloor_{r^{'} \in \mathcal{R}}^T \rfloor_{r \in \mathcal{R}}$ is the stacked self-interference channel, viewing all the FD-RU nodes as a single FD transceiver. It is worth mentioning that the values of $\kappa, \beta$ depend on the implemented SIC scheme and reflect the quality of the cancellation. For more discussions on the used distortion model please see \cite{DMBS:12, DMBSR:12, add_2, add_3, add_6, add_7, add_8, add_9, add_10}, and the references therein.       


\section{Achievable Secure Information Rate}
In this part, we express the achievable secure information rate, i.e., the information rate that can be transfered from (to) the core network to (from) the end-users while kept secure against the untrusted RUs and the eavesdroppers, as a function of transmission and compression strategies. In particular, the achievable rate expressions are obtained utilizing the following fundamental results. Firstly, we employ the proposed multivariate compression scheme proposed in \cite{MultivariateQuantization_1, MultivariateQuantization_2} and later used in \cite{8437192, park2017fronthaul} for similar purposes of preserving the information privacy. In particular to our work, the CU is able to correlate the quantization noise for different DL CU-RU fronthaul links, thereby enabling a mechanism for quantization noise covariance shaping and DL beamforming with the purpose of improving the information secrecy. Secondly, we assume Gaussian signal codewords as well as the Gaussian noise and distortion signal components, enabling the utilization of the Shannon's bound on the achievable information rate, see \cite{8709756} for a similar assumption set. And thirdly, we employ the results by \cite{liang2009compound} on the compound wiretap channel, indicating the achievable secure information rate among trusted entities in the presence of multiple untrusted entities, please also see~\cite{shamai_new, 8437192} for more elaborations and similar utilization of the aforementioned concepts. 

\subsubsection{Achievable UL/DL communication rate}
Assuming a sufficiently long coding block length as well as Gaussian distribution for all signal components, the achievable UL information rate, i.e., the achievable information rate among the $k$-th UL user and the CU can be expressed as 
\begin{align}    
R_{\text{ul},k} & = \text{log} \left| \sum_{i \in \mathcal{U}}  \ma{H}_{\text{ul}, i}  \tilde{\ma{F}}_{i} \ma{H}_{\text{ul}, i}^H + \ma{N}_{\text{ul}} + \ma{\Lambda} \left( \{\tilde{\ma{W}}_m \}, \ma{Q}_{\text{dl}} \right) + \ma{Q}_{\text{ul}} \right| \nonumber \\ 
& \;\;\;\;\;\;\;\;\;\; - \text{log} \left| \sum_{i \in \mathcal{U} \setminus k}  \ma{H}_{\text{ul}, i}  \tilde{\ma{F}}_{i} \ma{H}_{\text{ul}, i}^H + \ma{N}_{\text{ul}} + \ma{\Lambda} \left( \{\tilde{\ma{W}}_m \}, \ma{Q}_{\text{dl}} \right) + \ma{Q}_{\text{ul}}  \right|,\;\; \forall k \in \mathcal{U},
\end{align}      
incorporating the impact of residual self-interference, UL quantization, and inter-user interference. In the above expression, $\tilde{\ma{F}}_{m} := {\ma{F}}_{m} {\ma{F}}_{m}^H$ is the transmit UL covariance and $\ma{N}_{\text{ul}} = \left< N_{\text{ul},r}\ma{I}_{M_r} \right>_{r\in\mathcal{R}}$ is the stacked thermal noise covariance at the RUs. Similarly, the achievable DL information rate is written as
\begin{align} 
R_{\text{dl},m} &= \text{log}  \left| \sum_{i \in \mathcal{D}}  \ma{H}_{\text{dl},i} \tilde{\ma{W}}_i \ma{H}_{\text{dl},i}^H + \sum_{i \in \mathcal{U}}  \ma{H}_{\text{ud},im} \tilde{\ma{F}}_{i} \ma{H}_{\text{ud},im}^H + \ma{H}_{\text{dl},m} \ma{Q}_{\text{dl}}  \ma{H}_{\text{dl},m}^H  + N_{\text{dl},m}\ma{I}_{\tilde{M}_{m}} \right| \nonumber \\ 
& \;\;\;\;\;\;\;\; - \text{log}  \left| \sum_{i \in \mathcal{D} \setminus m}  \ma{H}_{\text{dl},i} \tilde{\ma{W}}_i \ma{H}_{\text{dl},i}^H + \sum_{i \in \mathcal{U}}  \ma{H}_{\text{ud},im} \tilde{\ma{F}}_{i} \ma{H}_{\text{ud,im}}^H  + \ma{H}_{\text{dl},m} \ma{Q}_{\text{dl}}  \ma{H}_{\text{dl},m}^H  + N_{\text{dl},m}\ma{I}_{\tilde{M}_{m}}  \right|,\;\; \forall m \in \mathcal{D},
\end{align}   
incorporating the impacts of DL quantization, thermal noise and co-channel interference.

\subsubsection{Pessimistic information leakage to RUs}
Assuming successive interference decoding and cancellation capability at the untrusted RUs for intercepting the UL streams~\cite{Schober_Sec8349956, Schober_Sec_7835110, 8709756}, an upper bound on the information leakage from the $k$-th UL user to the $r$-th RU can be expressed as 
\begin{align} \label{eq_leack_RU}
L^{\text{RU}}_{\text{ul},k,r} &= \text{log}  \left| \ma{H}_{\text{ul},kr} \tilde{\ma{F}}_{k} \ma{H}_{\text{ul},kr}^H +  \ma{S}_{\text{ul},r} \ma{\Lambda} \left( \{\tilde{\ma{W}}_m \}, \ma{Q}_{\text{dl}} \right) \ma{S}_{\text{ul},r} + \ma{H}_{\text{rr},r} \ma{Q}_{\text{dl}} \ma{H}_{\text{rr},r}^H  + N_{\text{ul},r}\ma{I}_{M_{r}} \right| \nonumber  \\
& \quad\quad - \text{log} \left| \ma{S}_{\text{ul},r} \ma{\Lambda} \left( \{\tilde{\ma{W}}_m \}, \ma{Q}_{\text{dl}} \right) \ma{S}_{\text{ul},r} + \ma{H}_{\text{rr},r} \ma{Q}_{\text{dl}} \ma{H}_{\text{rr},r}^H  + N_{\text{ul},r}\ma{I}_{M_{r}} \right|,
\end{align}
where $\ma{S}_{\text{ul},r}$ is the selection matrix defined in (\ref{eq_sel_matrix}). Please note that the above bound on the information leackage represents the pessimistic case where the untrusted RU may employ a non-linear processing strategy to decode the UL information, hence, considers the successive interference decoding and cancellation capability at the RU. 

Contrary to the UL information leakage where the RU could receive the related waveform only through the user-RU link, the RU may overhear the signal containing the DL waveforms through multiple paths. In particular, the RU may capture and store the DL waveform received from the CU through the fronthaul link, as well as through the inter-RU wireless channel from the RU-user communication. In order to jointly consider both reception paths, the stacked observation of the $m$-th DL user to the $r$-th RU is expressed as 
\begin{align} \label{ru_leack_sig_stack}
\tilde{\ma{y}}_{\text{leak},m,r} & = \left[\begin{array}{c} \ma{W}_{m,r} \\ \ma{H}_{\text{rr},r} \ma{W}_m \end{array}\right] \ma{s}_{\text{dl},m}  + \left[\begin{array}{c} \ma{q}_{\text{dl},r} \\ \ma{n}_{\text{ul},r} + \ma{H}_{\text{rr},r} \ma{q}_{\text{dl}} \end{array}\right]  \nonumber \\
 & = \underbrace{\left[\begin{array}{c} \ma{S}_{\text{dl},r} \\ \ma{H}_{\text{rr},r} \end{array}\right]}_{=: \ma{H}_{\text{eq},r}} \left( \ma{W}_m \ma{s}_{\text{dl},m} + \ma{q}_{\text{dl}}\right) + \left[\begin{array}{c} \ma{0}_{N_r\times 1} \\ \ma{n}_{\text{ul},r} \end{array}\right] := \ma{H}_{\text{eq},r} \left( \ma{W}_m \ma{s}_{\text{dl},m} + \ma{q}_{\text{dl}}\right) + \ma{n}_{\text{eq},r}, 
\end{align} 
where $\ma{H}_{\text{eq},r}$ denotes the effective combined channel among the $m$-th DL transmission and the $r$-th RU and $\ma{n}_{\text{eq},r} \sim \mathcal{CN} \left( \ma{0},  \ma{N}_{\text{eq},r}\right)$. Please note that similar to (\ref{eq_leack_RU}), the above expression considers the pessimistic situation that the untrusted node is capable of perfect SIC, e.g., via employing more sophisticated SIC by dedicating a larger processing power, for decoding/intercepting the information. A bound on the information leakage for the $m$-th DL user to the $r$-th RU can be hence expressed as 
\begin{align}
L_{\text{dl},m,r}^{\text{RU}} & = \text{log} \left| \ma{H}_{\text{eq},r} \tilde{\ma{W}}_m \ma{H}_{\text{eq},r}^H + \ma{H}_{\text{eq},r} \ma{Q}_{\text{dl}} \ma{H}_{\text{eq},r}^H +  \ma{N}_{\text{eq},r}  \right|- \text{log} \left| \ma{H}_{\text{eq},r} \ma{Q}_{\text{dl}} \ma{H}_{\text{eq},r}^H + \ma{N}_{\text{eq},r} \right|, 
\end{align} 
where $\ma{N}_{\text{eq},r}$ is the covariance of the stacked noise vector in (\ref{ru_leack_sig_stack}).

\subsubsection{Pessimistic information leakage to eavesdroppers}
Following a similar approach as for the RUs regarding the information leackage, we have
\begin{align}
L_{\text{ul},k,l}^{\text{Eve}} & = \text{log} \left| \ma{H}_{\text{ue},kl} \tilde{\ma{F}}_m \ma{H}_{\text{ue},kl}^H + \ma{H}_{\text{re},l} \ma{Q}_{\text{dl}} \ma{H}_{\text{re},l}^H +  {N}_{\text{e},l} \ma{I}  \right|- \text{log} \left| \ma{H}_{\text{re},l} \ma{Q}_{\text{dl}} \ma{H}_{\text{re},l}^H +  {N}_{\text{e},l} \ma{I} \right|, 
\end{align} 
and
\begin{align}
L_{\text{dl},m,l}^{\text{Eve}} & = \text{log} \left| \ma{H}_{\text{re},l} \tilde{\ma{W}}_m \ma{H}_{\text{re},m}^H + \ma{H}_{\text{re},l} \ma{Q}_{\text{dl}} \ma{H}_{\text{re},l}^H + {N}_{\text{e},l} \ma{I}  \right|- \text{log} \left| \ma{H}_{\text{re},l} \ma{Q}_{\text{dl}} \ma{H}_{\text{re},l}^H + {N}_{\text{e},l} \ma{I} \right|,                   
\end{align}                                                                  
respectively representing the information leakage from the UL and DL communications towards the eavesdroppers, where DL quantization noise is used as a friendly jamming signal towards the eavesdropper nodes to improve information secrecy.  

\subsubsection{Achievable Secrecy Rate}
Following \cite{liang2009compound}, the achievable secure information rate in the UL and in the DL can be hence expressed as
\begin{align}
R_{\text{sec-dl},m} & = \left\{ R_{\text{dl},m} -  \text{max} \; \left\{ \underset{r\in \mathcal{R}\setminus\mathcal{M}}{\text{max}} \;  L_{\text{dl},m,r}^{\text{RU}},\;\; \underset{l}{\text{max}} \;  L_{\text{dl},m,l}^{\text{Eve}} \right\}  \right\}^+, \\   
R_{\text{sec-ul},m} & = \left\{ R_{\text{ul},m} -  \text{max} \; \left\{ \underset{r\in \mathcal{R}\setminus\mathcal{M}}{\text{max}} \;  L_{\text{ul},m,r}^{\text{RU}},\;\; \underset{l}{\text{max}} \;  L_{\text{ul},m,l}^{\text{Eve}} \right\}  \right\}^+, 
\end{align} 
indicating the achievable communication rate which may not be decoded by any of the untrusted entities. Subsequently, the network sum secrecy rate is expressed as a function of the transmit and compression UL and DL covariance as  
\begin{align}
\text{WSSR} \left( \left\{ \tilde{\ma{W}}_m \right\}, \left\{ \tilde{\ma{F}}_m \right\}, \ma{Q}_{\text{dl}}, \ma{Q}_{\text{ul}} \right) = \sum_{m \in \mathcal{D}} w_m R_{\text{sec-dl},m} + \sum_{k \in \mathcal{U}} \bar{w}_k R_{\text{sec-ul},k}, 
\end{align}
where the weights ${w}_m, \bar{w}_k$ represent the significance of the obtained secrecy rate at each link and thereby incorporate specific service requirements to the design. 
\subsubsection{Fronthaul capacity constraints}
Employing the UL/DL transmit precoding and quantization strategies, the fronthaul load can be obtained as        
\begin{align}
F_{\text{dl},r} &= \text{log}  \left|  \ma{S}_{\text{dl},r} \left( \sum_{m \in \mathcal{D}} \tilde{\ma{W}}_m  + \ma{Q}_{\text{dl}} \right) \ma{S}_{\text{dl},r}^T  \right| - \text{log} \left|  \ma{S}_{{dl},r} \left( \sum_{m \in \mathcal{D}} \tilde{\ma{W}}_m   \right) \ma{S}_{\text{dl},r}^T  \right|, \\ 
F_{\text{ul},r} &= \text{log}  \left|    \ma{S}_{\text{ul},r} \left( \sum_{i \in \mathcal{U}}  \ma{H}_{\text{ul}, i}  \tilde{\ma{F}}_{i} \ma{H}_{\text{ul}, i}^H + \ma{N}_{\text{ul}} + \ma{\Lambda} \left( \{\tilde{\ma{W}}_m \}, \ma{Q}_{\text{dl}} \right) + \ma{H}_{\text{rr}} \left( \sum_{m \in \mathcal{D}} \tilde{\ma{W}}_m  + \ma{Q}_{\text{dl}} \right) \ma{H}_{\text{rr}}^H + \ma{Q}_{\text{ul}} \right)\ma{S}_{\text{ul},r}^T \right| \nonumber \\
& \;\;\;\;- \text{log}  \left| \ma{S}_{\text{ul},r} \left( \sum_{i \in \mathcal{U}}  \ma{H}_{\text{ul}, i}  \tilde{\ma{F}}_{i} \ma{H}_{\text{ul}, i}^H + \ma{N}_{\text{ul}} + \ma{\Lambda} \left( \{\tilde{\ma{W}}_m \}, \ma{Q}_{\text{dl}} \right) + \ma{H}_{\text{rr}} \left( \sum_{m \in \mathcal{D}} \tilde{\ma{W}}_m  + \ma{Q}_{\text{dl}} \right) \ma{H}_{\text{rr}}^H  \right)\ma{S}_{\text{ul},r}^T \right|,
\end{align}
respectively representing the required information rate of the DL and UL waveform transmissions over the fronthaul links with limited capacity. 
 
%
%
%
%
\section{Joint Transmission and Compression Optimization: An SDR-GIA Approach}
This is the purpose of this part to optimize the transmission and compression strategies through the network. In particular, the covariance of the DL and UL transmissions, as well as the DL and UL quantization strategies must be chosen with the goal of maximizing the achievable WSSR. The corresponding optimization problem can be hence formulated as
\begin{subequations}  \label{P_0}
\begin{align} \label{Opt_Original}
\underset{ \{\tilde{\ma{W}}_m\}, \{\tilde{\ma{F}}_k\} \ma{Q}_{\text{dl}}, \ma{Q}_{\text{ul}}}{\text{max}} \;\; &  \text{WSSR} \\
\text {s.t.} \;\;\; & F_{\text{ul},r} \leq C_{\text{ul},r}, \;\; \forall r \in \mathcal{R}, \\
& F_{\text{dl},r} \leq C_{\text{dl},r}, \;\; \forall r \in \mathcal{R}, \\
& \text{tr} \left( \ma{S}_{\text{dl},r} \left( \sum_{m \in \mathcal{D}} \tilde{\ma{W}}_m  + \ma{Q}_{\text{dl}} \right) \ma{S}_{\text{dl},r}^T \right) \leq P_{\text{dl},r}, \;\; \forall r \in \mathcal{R}, \\
& \text{tr} \left( \tilde{\ma{F}}_k \right)  \leq P_{\text{ul},k}, \;\; \forall k \in \mathcal{U},\\
& \tilde{\ma{W}}_m, \tilde{\ma{F}}_k,\ma{Q}_{\text{dl}}, \ma{Q}_{\text{ul}} \succeq \ma{0}, \label{semidefinite_consts}\\
& \text{rank} \left(\tilde{\ma{W}}_m\right) \leq d_m. \label{rank_consts} 
\end{align}   
\end{subequations}
In the above problem, (\ref{P_0}b)-(\ref{P_0}c) represent the constraint on fronthaul load and (\ref{P_0}d)-(\ref{P_0}e) represent the power constraints. The constraints (\ref{P_0}f) and (\ref{P_0}g) respectively impose the positive semi-definiteness and low-rank structure, which are necessary to obtain a feasible and constructible transmit covariance. It can be observed that the above problem is not mathematically tractable, due to the non-linear and non-convex objective as well as the non-convex constraint sets. In order to obtain a tractable form, the epigraph form of (\ref{P_0}) is formulated as 
\begin{subequations}  \label{P_1}
\begin{align} \label{Opt_Original_1}
\underset{ \{\tilde{\ma{W}}_m\}, \{\tilde{\ma{F}}_k\}, \{\zeta_m, \bar{\zeta}_m \}, \{\gamma_k, \bar{\gamma}_k \},  \ma{Q}_{\text{dl}}, \ma{Q}_{\text{ul}} }{\text{max}} \;\; & \sum_{m \in \mathcal{D}} {w}_m \left(\bar{\zeta}_m - \zeta_m \right) + \sum_{k \in \mathcal{U}} \bar{w}_k \left(\bar{\gamma}_k - \gamma_k  \right)  \\
\text {s.t.} \;\;\; 
& R_{\text{dl},m} \geq \bar{\zeta}_m , \;\; \forall m\in \mathcal{D},\\
& R_{\text{ul},k} \geq \bar{\gamma}_k , \;\; \forall k\in \mathcal{U},\\
& L_{\text{dl},m,r}^{\text{RU}} \leq \zeta_m , \;\; \forall m\in \mathcal{D}, \;r\in \mathcal{R}\setminus \mathcal{M},  \\
& L_{\text{dl},m,l}^{\text{Eve}} \leq \zeta_m , \;\; \forall m\in \mathcal{D}, \;l\in \mathcal{E},  \\
& L_{\text{ul},k,r}^{\text{RU}} \leq \gamma_k ,  \;\; \forall k\in \mathcal{U}, \;r\in \mathcal{R}\setminus \mathcal{M},  \\
& L_{\text{ul},k,l}^{\text{Eve}} \leq \gamma_k, \;\; \forall k\in \mathcal{U}, \;r\in \mathcal{E}, \\
& F_{\text{ul},r} \leq C_{\text{ul},r}, \;\; \forall r \in \mathcal{R}, \\
& F_{\text{dl},r} \leq C_{\text{dl},r}, \;\; \forall r \in \mathcal{R}, \\
& \text{(\ref{P_0}d)-(\ref{P_0}g)},
\end{align}            
\end{subequations}     
where (\ref{P_1}b)-(\ref{P_1}g) define the epigraph form of the various rate expressions and $\bar{\gamma}_m,\bar{\zeta}_k, \zeta_m, \gamma_k \in \mathbb{R}$ are the introduced auxiliary variables. Please note that at the optimality of (\ref{P_0}), the non-negativeness operator $\{.\}^+$ has no effect, and hence it is dropped thereafter in formulating the optimization objective\footnote{This statement follows, similar to that of \cite{8709756}, from the observation that if at the optimality of (\ref{P_0}) any of the expressions $R_{\text{ul},m} -  \text{max} \; \left\{ \underset{r\in \mathcal{R}\setminus\mathcal{M}}{\text{max}} \;  L_{\text{ul},m,r}^{\text{RU}},\;\; \underset{l}{\text{max}} \;  L_{\text{ul},m,l}^{\text{Eve}} \right\}$ or $R_{\text{dl},m} -  \text{max} \; \left\{ \underset{r\in \mathcal{R}\setminus\mathcal{M}}{\text{max}} \;  L_{\text{dl},m,r}^{\text{RU}},\;\; \underset{l}{\text{max}} \;  L_{\text{dl},m,l}^{\text{Eve}} \right\}$ hold a negative value, the transmit covariance $\tilde{\ma{W}}_m $ or $\tilde{\ma{F}}_k$ can be put to zero to improve the negative value (and equalizes it to zero) which leads to contradiction.}. Please note that the above problem is still not tractable, due to the non-convex feasible set. In order to proceed, we first relax the non-convex rank constraint in (\ref{P_0}g), employing the SDR framework. Furthermore, we recognize that the non-convex constraints (\ref{P_1}b)-(\ref{P_1}g) and (\ref{P_0}b)-(\ref{P_0}c) can be all presented via smooth difference-of-convex (DC) functions, thereby enabling application of the general inner approximation (GIA) framework, with convergence to a solution satisfying Karush–Kuhn–Tucker (KKT) optimality conditions. In particular, let the set $\mathcal{V}$ be defined as 
\begin{align}
\mathcal{V} := \left\{\{\tilde{\ma{W}}_m\}, \{\tilde{\ma{F}}_k\}, \{\zeta_m, \bar{\zeta}_m \}, \{\gamma_k, \bar{\gamma}_k \},  \ma{Q}_{\text{dl}}, \ma{Q}_{\text{ul}} \right\},
\end{align}
representing the problem variable set. By employing Taylor's approximation on the concave parts of the DC expressions, the optimization problem (\ref{P_1}) is approximated at the given point $\mathcal{V}_0$ as   
                                      
\begin{subequations}  \label{P_2}     
\begin{align} \label{Opt_Original_P_2}
\underset{ \mathcal{V}}{\text{max}} \;\; & \sum_{m \in \mathcal{D}} {w}_m \left( \bar{\zeta}_m - \zeta_m \right) + \sum_{k \in \mathcal{U}} \bar{w}_k \left( \bar{\gamma}_k - \gamma_k \right)  \\
\text {s.t.} \;\;\;  
& \tilde{R}_{\text{dl},m} \left( \mathcal{V},\mathcal{V}_0 \right) \geq \bar{\gamma}_m , \;\; \forall m\in \mathcal{D},\\
& \tilde{R}_{\text{ul},k} \left( \mathcal{V},\mathcal{V}_0 \right) \geq \bar{\zeta}_k , \;\; \forall k\in \mathcal{U},\\
& \tilde{L}_{\text{dl},m,r}^{\text{RU}} \left( \mathcal{V},\mathcal{V}_0 \right) \leq \zeta_m , \;\; \forall m\in \mathcal{D}, \;r\in \mathcal{R}\setminus \mathcal{M},  \\
& \tilde{L}_{\text{dl},m,l}^{\text{Eve}} \left( \mathcal{V},\mathcal{V}_0 \right) \leq \zeta_m , \;\; \forall m\in \mathcal{D}, \;l\in \mathcal{E},  \\
& \tilde{L}_{\text{ul},k,r}^{\text{RU}} \left( \mathcal{V},\mathcal{V}_0 \right) \leq \gamma_k ,  \;\; \forall k\in \mathcal{U}, \;r\in \mathcal{R}\setminus \mathcal{M},  \\
& \tilde{L}_{\text{ul},k,l}^{\text{Eve}} \left( \mathcal{V},\mathcal{V}_0 \right) \leq \gamma_k, \;\; \forall k\in \mathcal{U}, \;l\in \mathcal{E}, \\
& \tilde{F}_{\text{ul},r} \left( \mathcal{V},\mathcal{V}_0 \right) \leq C_{\text{ul},r}, \;\; \forall r \in \mathcal{R}, \\
& \tilde{F}_{\text{dl},r} \left( \mathcal{V},\mathcal{V}_0 \right) \leq C_{\text{dl},r}, \;\; \forall r \in \mathcal{R}, \\
& \text{(\ref{P_0}d)-(\ref{P_0}f)},
\end{align}
\end{subequations}
where the expressions $\tilde{R}_{\text{dl},m}, \tilde{R}_{\text{ul},k}$ and $\tilde{L}_{\text{dl},m,r}^{\text{RU}}, \tilde{L}_{\text{dl},m,l}^{\text{Eve}}, \tilde{L}_{\text{ul},k,r}^{\text{RU}} , \tilde{L}_{\text{ul},k,l}^{\text{RU}}, \tilde{F}_{\text{ul},r}, \tilde{F}_{\text{dl},r}$, respectively, are the constructed lower and upper bounds defined in (\ref{eq_approximations}a)-(\ref{eq_approximations}g), constituting the inner convex approximations corresponding to the constraints (\ref{P_2}b)-(\ref{P_2}i). 
Please note that in the approximations (\ref{eq_approximations}a)-(\ref{eq_approximations}g), the function 
\begin{align} \label{Taylor_log}
\varphi (\ma{X}, \ma{X}_0):= \text{log}\left|\ma{X}_0\right| + \frac{1}{\text{ln}(2)} \left( \text{tr}\left( \left(\ma{X}_0\right)^{-1}  \left( \ma{X} - \ma{X}_0 \right) \right) \right), 
\end{align}
obtains an affine upper bound of the concave logarithmic function $\text{log}\left|\ma{X}\right|$ at the point $\ma{X}_0$ via Taylor's approximation and thereby constitutes a tight (at $\ma{X}_0$) and global affine upper-approximator to the concave expressions. In the following, we define an iterative algorithm to solve (\ref{P_0}) employing the approximation (\ref{P_2}). Please note that the problem (\ref{P_1}) is an instance of smooth difference-of-convex programs, complying with the GIA framework, presented in \cite{marks1978technical}. Furthermore, the obtained approximated problem (\ref{P_2}) is a convex program and can be solved to the optimality via e.g., interior point methods~\cite{BV:04, bertesekas1999nonlinear}. The iterations of inner approximation and consequently solving (\ref{P_2}) will be continued until a stable solution is obtained, please see Algorithm~1 for the detailed procedures. 

\subsection{Convergence}
Algorithm~\ref{alg_GIA} converges to a solution satisfying the KKT optimality conditions of the equivalent problem (\ref{P_1}) with relaxed rank constraints. In order to observe this, we recall that the approximations (\ref{eq_approximations}a)-(\ref{eq_approximations}g) are obtained utilizing the Taylor's approximation on a smooth concave function, i.e., (\ref{Taylor_log}). In particular, we observe the following properties: \emph{i)} $\text{log}\left(\ma{X}_0\right) = \varphi\left(\ma{X}_0,\ma{X}_0\right)$, i.e., the tightness property, \emph{ii)} $\text{log}\left(\ma{X}\right) \leq \varphi\left(\ma{X},\ma{X}_0\right), \; \forall \ma{X}$, globally upper-bound property, and \emph{iii)} $\partial\text{log}\left(\ma{X}\right)/\partial\ma{X} = \partial \varphi \left(\ma{X},\ma{X}_0\right)/\partial\ma{X}\big|_{\ma{X}=\ma{X}_0}$, property of shared slope at the point of approximation. 
Consequently, the constructed approximations in (\ref{eq_approximations}a)-(\ref{eq_approximations}g) also satisfy the properties stated in \cite[Theorem~1]{marks1978technical}. This concludes the convergence of the sequence generated by (\ref{P_2}) to a KKT point of (\ref{P_1}) with a relaxed rank constraint.
\subsection{Rank reduction}
Due to the nature of the SDP program in (\ref{P_2}), the obtained solutions for the DL transmit covariance matrices, i.e., $\tilde{\ma{W}}_m, \; \forall m \in \mathcal{D}$, do not necessarily satisfy the rank constraint which is imposed initially by (\ref{P_0}g). Please note that the transmit covariance of a higher rank can not be realized via standard linear transmit/receive signal processing, due to the limited number of antennas at the receiver. In order to obtain a feasible solution, Gaussian randomization method is widely used, where the rank-reduced solution is chosen from a set of randomly generated solution candidates. Nevertheless, in particular to our system, the aforementioned method leads to a necessary re-adjustment in the studied problem, which leads to a high computational complexity. In order to resolve this issue, we propose an iterative rank-reduction procedure, where the constraint (\ref{P_0}g) is satisfied by iteratively cutting the problem feasible space. The implemented rank-reduction procedures are summarized in the following: 
\subsubsection{Gaussian Randomization (GR)}\label{sec:gaussian_randomization}
Let $\tilde{\ma{W}}_m^{\star}$ be the obtained DL transmit covariance from (\ref{P_2}), with the singular value decomposition as $\tilde{\ma{W}}_m^{\star} = \ma{U}_m \ma{\Sigma}_m \ma{U}_m^H, \; \forall m$. For each instance of the GR, we generate random matrices  $\ma{X}^{(\ell)} \leftarrow \mathcal{CN} \left( \ma{0}_{ \sum_{r \in \mathcal{R}} N_r \times d_m}, \ma{I}_{\sum_{r \in \mathcal{R}} N_r}  \right)$. The resulting rank-constrained matrix is then calculated as $\ma{W}_m^{(\ell)} \leftarrow \ma{U}_m \left( \ma{\Sigma}_m \right)^{\frac{1}{2}} \ma{X}^{(\ell)},\; \forall m$, satisfying the intended rank constraint (\ref{P_0}g).  Please note that although the random generation is guaranteed to satisfy the rank constraint, it may render the other problem constraints (\ref{P_1}b)-(\ref{P_1}g) infeasible. In this regard, a scalar adjustment is required on the obtained low-rank solutions, by continuing the iterates of (\ref{P_2}) until convergence over the scalar variable set $\{\theta_m\}, \{\zeta_m, \bar{\zeta}_m \}, \{\gamma_k, \bar{\gamma}_k\}$, where $\theta_m$ is the scaling factor adopted for $\tilde{\ma{W}}_m^{(\ell)}$. The eventual choice of $\tilde{\ma{W}}_m^{\star}$ is then chosen as the best-performing solution among the recovered feasible candidates $\ma{W}_m^{(\ell)}$ via GR. 
\subsubsection{Iterative Reduction Method}\label{sec:iterative_reduction}
It is observed that the well-known randomization method incurs a high computational complexity for the problem at hand, due to the necessary re-adjustments which need to be repeated as a separate optimization problem for \textit{each instance} of the random generation. In order to obtain an efficient solution, we propose an iterative method where the feasible space associated with the matrices $\tilde{\ma{W}}_m$ is sequentially reduced in order to comply with the rank constraint. In this regard, when a transmit DL covariance exceeds the constructible matrix rank, we impose a new linear constraint on $\tilde{\ma{W}}_m$ with the role of eliminating its permissible column space in the least effective singular mode, thereby limiting the feasible column space of $\tilde{\ma{W}}_m$ and the resulting matrix rank in the subsequent iterations. The updated problem is expressed as 
\begin{subequations}  \label{P_3}
\begin{align} \label{Opt_Original_P_3}
\underset{ \mathcal{V}}{\text{max}} \;\; & \sum_{m \in \mathcal{D}} {w}_m \left(\bar{\zeta}_m - \zeta_m \right) + \sum_{k \in \mathcal{U}} \bar{w}_k \left(\bar{\gamma}_k - \gamma_k  \right)     \\
\text {s.t.} \;\;\;  
&  \text{tr}\left( \tilde{\ma{W}}_m \ma{J}_m \right) = 0, \forall m \in \mathcal{D}, \\
& \text{(\ref{P_2}b)-(\ref{P_2}i),~(\ref{P_0}d)-(\ref{P_0}f)},
\end{align}
\end{subequations}
where $\ma{J}_m$ contains the column space which is reduced from the feasible space of $\tilde{\ma{W}}_m$. In the first iteration, we employ the initialization $\ma{J}_m = \ma{0}$ which corresponds to no constraint on $\tilde{\ma{W}}_m$. For every stationary point of the problem (\ref{P_3}), the matrices $\ma{J}_m$ are updated to prohibit the least effective eigenmodes for the matrices where a rank violation occurs, thereby reducing the permissible maximum rank. In order to establish this, 
Let $\tilde{\ma{W}}_m^{\star}$ be the obtained DL transmit covariance from (\ref{P_3}), with the singular value decomposition as $\tilde{\ma{W}}_m^{\star} = \ma{U}_m \ma{\Sigma}_m \ma{U}_m^H, \; \forall m $. Furthermore, let 
\begin{align}
\ma{U}_m = \left[ \ma{u}_1,\cdots \ma{u}_{d_m}, \ma{u}_{{d_m}+1}, \cdots,  \ma{u}_{\sum_{r \in \mathcal{R}} N_r} \right]
\end{align}

\begin{algorithm}[!t]
 \small{	\begin{algorithmic}[1]
\State{$\text{Initialize}\; \mathcal{V}^{[0]},\; \ma{J}_m \leftarrow \ma{0}, \; \forall m \in \mathcal{D}, \; i \leftarrow 0,$}
\Repeat   
\State{$i \leftarrow i + 1,$}
\State{$\mathcal{V}^{[i]} \leftarrow \text{solve}\;(\ref{P_3}),$} 
\If{\text{Convergence of (\ref{P_3})~\textbf{AND}~Case~1 }}
\Comment{Guassian~Randomization}
\State{$\tilde{\ma{W}}_m, \forall m \in \mathcal{D},  \leftarrow$ Subsection~\ref{sec:gaussian_randomization},}
\State{\textbf{break}}
\EndIf
\If{\text{Convergence of (\ref{P_3})~\textbf{AND}~Case~2 }}
\Comment{Iterative~Reduction}
\State{$ \ma{J}_m, \forall m \in \mathcal{D}, \leftarrow$ Subsection~\ref{sec:iterative_reduction},}
\EndIf
\Until{Convergences, or maximum number of iterations reached }
\State{\Return{$\{\tilde{\ma{W}}_m\},\{\tilde{\ma{F}}_k\}, \ma{Q}_{\text{ul}}, \ma{Q}_{\text{dl}}$}}
 \end{algorithmic} } 
 \caption{{GIA-SDR based algorithm for solving (\ref{P_0}). $\epsilon$ determines the stability threshold.} }  \label{alg_GIA}  
\end{algorithm}
ordered in a descending manner according to the singular values in $\ma{\Sigma}_m$. The update of $\ma{J}_m$ is done as following
\begin{align}
\ma{J}_m = \left\{ \begin{array}{cc} \ma{J}_m,  &   \text{rank}\left( \tilde{\ma{W}}_m^{\star} \right) \leq d_m \\ \ma{J}_m + \ma{u}_{{d_m}+1}\ma{u}_{{d_m}+1}^H &  \text{rank}\left( \tilde{\ma{W}}_m^{\star} \right) > d_m \end{array}, \right.  \;\;  \forall m \in \mathcal{D}. 
\end{align}
The updates on $\ma{J}_m$ and the iterations of the optimization problem (\ref{P_3}) are continued until convergence of (\ref{P_3}) is achieved such that (\ref{P_0}g) is satisfied, please see Case~$2$ in Algorithm~1 for the algorithmic procedures.

\begin{figure*}[!h]
\scriptsize
\begin{subequations}\label{eq_approximations}
\begin{align}
\tilde{R}_{\text{ul},k} & = \text{log} \left| \sum_{i \in \mathcal{U}} \ma{H}_{\text{ul}, i}  \tilde{\ma{F}}_{i} \ma{H}_{\text{ul}, i}^H + \ma{N}_{\text{ul}} + \ma{\Lambda} \left( \{\tilde{\ma{W}}_m \}, \ma{Q}_{\text{dl}} \right) + \ma{Q}_{\text{ul}} \right| - \nonumber \\ 
& \hspace{-3mm} \varphi \Bigg{(} \sum_{i \in \mathcal{U} \setminus k}  \ma{H}_{\text{ul}, i}  \tilde{\ma{F}}_{i} \ma{H}_{\text{ul}, i}^H + \ma{N}_{\text{ul}} + \ma{\Lambda} \left( \{\tilde{\ma{W}}_m \}, \ma{Q}_{\text{dl}} \right) + \ma{Q}_{\text{ul}} , 
\sum_{i \in \mathcal{U} \setminus k}  \ma{H}_{\text{ul}, i}  \tilde{\ma{F}}_{i}^{0} \ma{H}_{\text{ul}, i}^H + \ma{N}_{\text{ul}} + \ma{\Lambda} \left( \{\tilde{\ma{W}}_m^{0} \}, \ma{Q}_{\text{dl}}^{0} \right) + \ma{Q}_{\text{ul}}^{0}
  \Bigg{)}, \\
\tilde{R}_{\text{dl},m} &= \text{log}  \left| \sum_{i \in \mathcal{D}}  \ma{H}_{\text{dl},i} \tilde{\ma{W}}_i \ma{H}_{\text{dl},i}^H + \sum_{i \in \mathcal{U}}  \ma{H}_{\text{ud},im} \tilde{\ma{F}}_{i} \ma{H}_{\text{ud},im}^H + \ma{H}_{\text{dl},m} \ma{Q}_{\text{dl}}  \ma{H}_{\text{dl},m}^H  + N_{\text{dl},m}\ma{I}_{\tilde{M}_{m}} \right| \nonumber \\ 
& \;\; -  \varphi \Bigg{(} \sum_{i \in \mathcal{D} \setminus m}  \ma{H}_{\text{dl},i} \tilde{\ma{W}}_i \ma{H}_{\text{dl},i}^H + \sum_{i \in \mathcal{U}}  \ma{H}_{\text{ud},im} \tilde{\ma{F}}_{i} \ma{H}_{\text{ud},im}^H  + \ma{H}_{\text{dl},m} \ma{Q}_{\text{dl}}  \ma{H}_{\text{dl},m}^H  + N_{\text{dl},m}\ma{I}_{\tilde{M}_{m}}, \nonumber \\ 
&\;\;\;\;\;\;\;\;\;\;\;\; \sum_{i \in \mathcal{D} \setminus m}  \ma{H}_{\text{dl},i} \tilde{\ma{W}}_i^{0} \ma{H}_{\text{dl},i}^H + \sum_{i \in \mathcal{U}}  \ma{H}_{\text{ud},im} \tilde{\ma{F}}_{i}^{0} \ma{H}_{\text{ud},im}^H  + \ma{H}_{\text{dl},m} \ma{Q}_{\text{dl}}^{0}  \ma{H}_{\text{dl},m}^H  + N_{\text{dl},m}\ma{I}_{\tilde{M}_{m}} \Bigg{)}, \\
\tilde{L}^{\text{RU}}_{\text{ul},k,r} &= \varphi \Bigg{(} \ma{H}_{\text{ul},kr} \tilde{\ma{F}}_{k} \ma{H}_{\text{ul},kr}^H +  \ma{S}_{\text{ul},r} \ma{\Lambda} \left( \{\tilde{\ma{W}}_m \}, \ma{Q}_{\text{dl}} \right) \ma{S}_{\text{ul},r}^T + \ma{H}_{\text{rr},r} \ma{Q}_{\text{dl}} \ma{H}_{\text{rr},r}^H  + N_{\text{ul},r}\ma{I}_{M_{r}}, \nonumber  \\
& \;\;\;\;\;\;\;\;\;\;\;\;\;\;\;\;\;\;\;\;\;\; \ma{H}_{\text{ul},kr} \tilde{\ma{F}}_{k}^{0} \ma{H}_{\text{ul},kr}^H +  \ma{S}_{\text{ul},r} \ma{\Lambda} \left( \{\tilde{\ma{W}}_m^{0} \}, \ma{Q}_{\text{dl}}^{0} \right) \ma{S}_{\text{ul},r}^T + \ma{H}_{\text{rr},r} \ma{Q}_{\text{dl}}^{0} \ma{H}_{\text{rr},r}^H  + N_{\text{ul},r}\ma{I}_{M_{r}} \Bigg{)} \nonumber \\
& \;\;\;\;\;\;\;\;\;\;\;\;\;\;\;\;\;\;\;\;\;\;\;\;\;\;\;\;\;\;\;\;\;\;\;\;\;\;\;\;\;\; - \text{log} \left| \ma{S}_{\text{ul},r} \ma{\Lambda} \left( \{\tilde{\ma{W}}_m \}, \ma{Q}_{\text{dl}} \right) \ma{S}_{\text{ul},r}^T + \ma{H}_{\text{rr},r} \ma{Q}_{\text{dl}} \ma{H}_{\text{rr},r}^H  + N_{\text{ul},r}\ma{I}_{M_{r}} \right|, \\
\tilde{L}_{\text{dl},m,r}^{\text{RU}} & = \varphi \Bigg{(} \ma{H}_{\text{eq},r} \tilde{\ma{W}}_m \ma{H}_{\text{eq},r}^H + \ma{H}_{\text{eq},r} \ma{Q}_{\text{dl}} \ma{H}_{\text{eq},r}^H +  \ma{N}_{\text{eq},r} , \ma{H}_{\text{eq},r} \tilde{\ma{W}}_m^{0} \ma{H}_{\text{eq},r}^H + \ma{H}_{\text{eq},r} \ma{Q}_{\text{dl}}^{0} \ma{H}_{\text{eq},r}^H +  \ma{N}_{\text{eq},r}   \Bigg{)}  \nonumber \\ 
& \hspace{55mm}\;\;\;\;\;\;\;\;\;\;\;\;\;\;\;\;\;\;\;\;\;\;\;\;\;\;\;\;\;\;\;\;\;\;\;\; - \text{log} \left| \ma{H}_{\text{eq},r} \ma{Q}_{\text{dl}} \ma{H}_{\text{eq},r}^H + \ma{N}_{\text{eq},r} \right|, \\         
\tilde{L}_{\text{ul},k,l}^{\text{Eve}} & = \varphi \Bigg{(} \ma{H}_{\text{ue},kl} \tilde{\ma{F}}_m \ma{H}_{\text{ue},kl}^H + \ma{H}_{\text{re},l} \ma{Q}_{\text{dl}} \ma{H}_{\text{re},l}^H +  {N}_{\text{e},l} \ma{I},  \;  \ma{H}_{\text{ue},kl} \tilde{\ma{F}}_m^{0} \ma{H}_{\text{ue},kl}^H + \ma{H}_{\text{re},l} \ma{Q}_{\text{dl}}^{0} \ma{H}_{\text{re},l}^H +  {N}_{\text{e},l} \ma{I} \Bigg{)}  \nonumber \\ 
& \hspace{55mm} \;\;\;\;\;\;\;\;\;\;\;\;\;\;\;\;\;\;\;\;\;\;\;\;\;\;\;\;\;\;\;\;\;\;\;\;\;\;- \text{log} \left| \ma{H}_{\text{re},l} \ma{Q}_{\text{dl}} \ma{H}_{\text{re},l}^H +  {N}_{\text{e},l} \ma{I} \right|, \\
\tilde{L}_{\text{dl},m,l}^{\text{Eve}} & = \varphi \Bigg{(} \ma{H}_{\text{re},l} \tilde{\ma{W}}_m \ma{H}_{\text{re},m}^H + \ma{H}_{\text{re},l} \ma{Q}_{\text{dl}} \ma{H}_{\text{re},l}^H + {N}_{\text{e},l} \ma{I} ,  \ma{H}_{\text{re},l} \tilde{\ma{W}}_m^{0} \ma{H}_{\text{re},m}^H + \ma{H}_{\text{re},l} \ma{Q}_{\text{dl}}^{0} \ma{H}_{\text{re},l}^H + {N}_{\text{e},l} \ma{I} \Bigg{)}\nonumber \\ 
& \hspace{55mm} \;\;\;\;\;\;\;\;\;\;\;\;\;\;\;\;\;\;\;\;\;\;\;\;\;\;\;\;\;\;\;\;\;\;\;\;\;\; - \text{log} \left| \ma{H}_{\text{re},l} \ma{Q}_{\text{dl}} \ma{H}_{\text{re},l}^H + {N}_{\text{e},l} \ma{I} \right|,   \\
\tilde{F}_{\text{dl},r} &= \varphi \Bigg{(}  \ma{S}_{\text{dl},r} \left( \sum_{m \in \mathcal{D}} \tilde{\ma{W}}_m  + \ma{Q}_{\text{dl}} \right) \ma{S}_{\text{dl},r}^T , \ma{S}_{\text{dl},r} \left( \sum_{m \in \mathcal{D}} \tilde{\ma{W}}_m^{0}  + \ma{Q}_{\text{dl}}^{0} \right) \ma{S}_{\text{dl},r}^T \Bigg{)}  - \text{log} \left|  \ma{S}_{{dl},r} \left( \sum_{m \in \mathcal{D}} \tilde{\ma{W}}_m   \right) \ma{S}_{\text{dl},r}^T  \right|, \\ 
\tilde{F}_{\text{ul},r} &= \varphi \Bigg{(}    \ma{S}_{\text{ul},r} \left( \sum_{i \in \mathcal{U}}  \ma{H}_{\text{ul}, i}  \tilde{\ma{F}}_{i} \ma{H}_{\text{ul}, i}^H + \ma{N}_{\text{ul}} + \ma{\Lambda} \left( \{\tilde{\ma{W}}_m \}, \ma{Q}_{\text{dl}} \right) + \ma{H}_{\text{rr}} \left( \sum_{m \in \mathcal{D}} \tilde{\ma{W}}_m  + \ma{Q}_{\text{dl}} \right) \ma{H}_{\text{rr}}^H + \ma{Q}_{\text{ul}} \right)\ma{S}_{\text{ul},r}^T,  \nonumber \\
& \;\;\ma{S}_{\text{ul},r} \left( \sum_{i \in \mathcal{U}}  \ma{H}_{\text{ul}, i}  \tilde{\ma{F}}_{i}^{0} \ma{H}_{\text{ul}, i}^H + \ma{N}_{\text{ul}} + \ma{\Lambda} \left( \{\tilde{\ma{W}}_m^{0} \}, \ma{Q}_{\text{dl}}^{0} \right) + \ma{H}_{\text{rr}} \left( \sum_{m \in \mathcal{D}} \tilde{\ma{W}}_m^{0}  + \ma{Q}_{\text{dl}}^{0} \right) \ma{H}_{\text{rr}}^H + \ma{Q}_{\text{ul}}^{0} \right)\ma{S}_{\text{ul},r}^T \Bigg{)} - \nonumber \\
& \;\;\text{log}  \left| \ma{S}_{\text{ul},r} \left( \sum_{i \in \mathcal{U}}  \ma{H}_{\text{ul}, i}  \tilde{\ma{F}}_{i} \ma{H}_{\text{ul}, i}^H + \ma{N}_{\text{ul}} + \ma{\Lambda} \left( \{\tilde{\ma{W}}_m \}, \ma{Q}_{\text{dl}} \right) + \ma{H}_{\text{rr}} \left( \sum_{m \in \mathcal{D}} \tilde{\ma{W}}_m  + \ma{Q}_{\text{dl}} \right) \ma{H}_{\text{rr}}^H  \right)\ma{S}_{\text{ul},r}^T \right|, \\
& \hspace{80mm} \forall l \in \mathcal{E},\;\; \forall k \in \mathcal{U}, \;\; \forall m \in \mathcal{D}, \;\; \forall r \in \mathcal{R}.
\end{align}        
\end{subequations} 
\vspace*{-0mm}     
\end{figure*}
 
\begin{figure}[!t]  
\hspace{0cm}\subfigure[Network Setup]{\includegraphics[height =  0.45\columnwidth, width = 0.5\columnwidth]{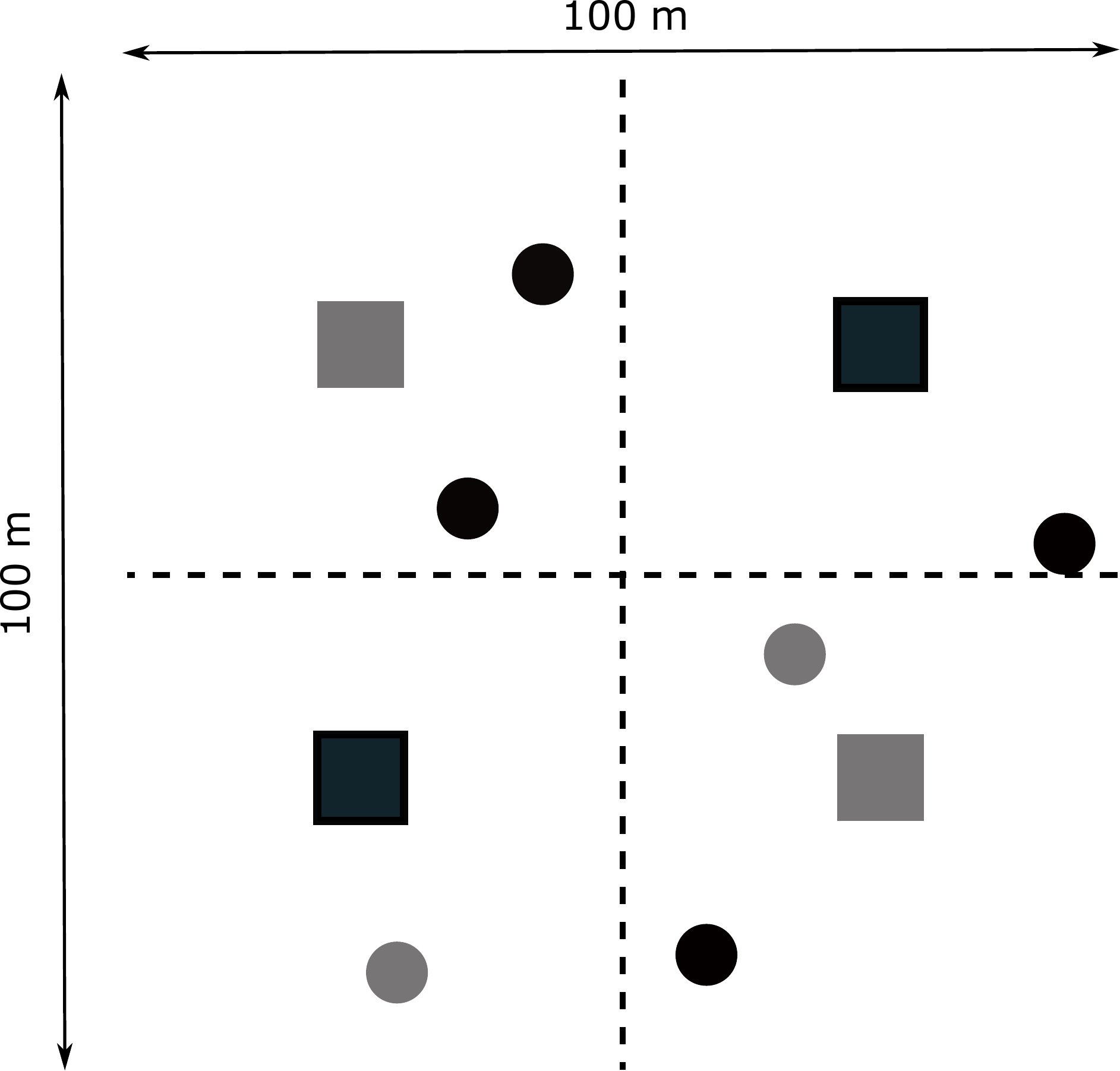} }
\hspace{0cm}\subfigure[Alg. Convergence]{\includegraphics[height =  0.45\columnwidth, width = 0.5\columnwidth]{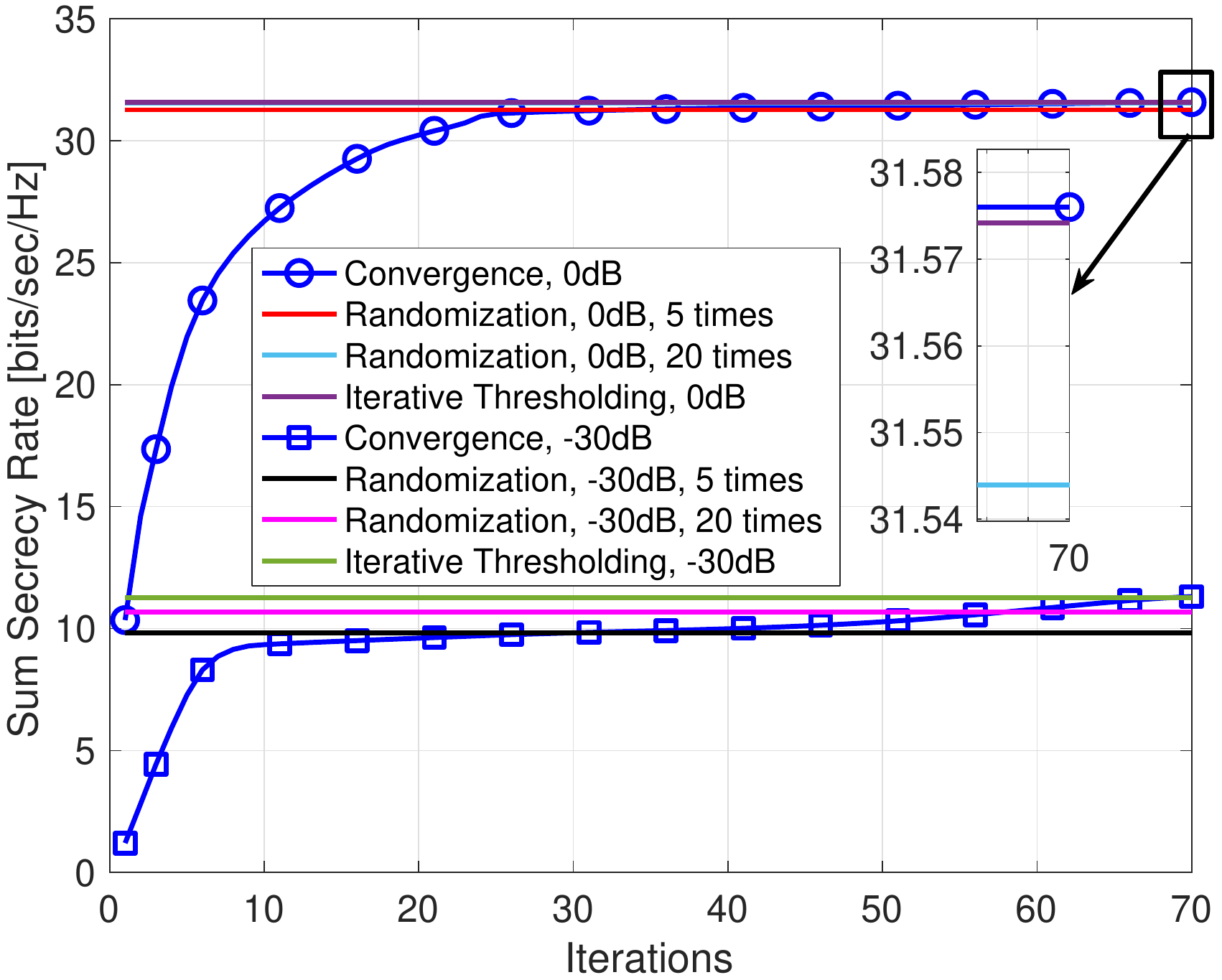} }   
    \caption{\textit{\textbf{(a)}}: Simulated network setup. The solid black (gray) squares represent the deployed trusted (untrusted) RU nodes at the center of the sub-squares. The users and the potential eavesdropper nodes are distributed randomly within the cell area where the solid black (gray) circles respectively represent the users and the potential eavesdroppers.~\textit{\textbf{(b)}}: Average convergence behavior of Algorithm~1.} \label{fig_general}
\end{figure}

\begin{figure}[!t]  
\hspace{0cm}\subfigure[Sum Secrecy Rate vs. $P_{\text{bud}}$]{\includegraphics[height =  0.45\columnwidth, width = 0.5\columnwidth]{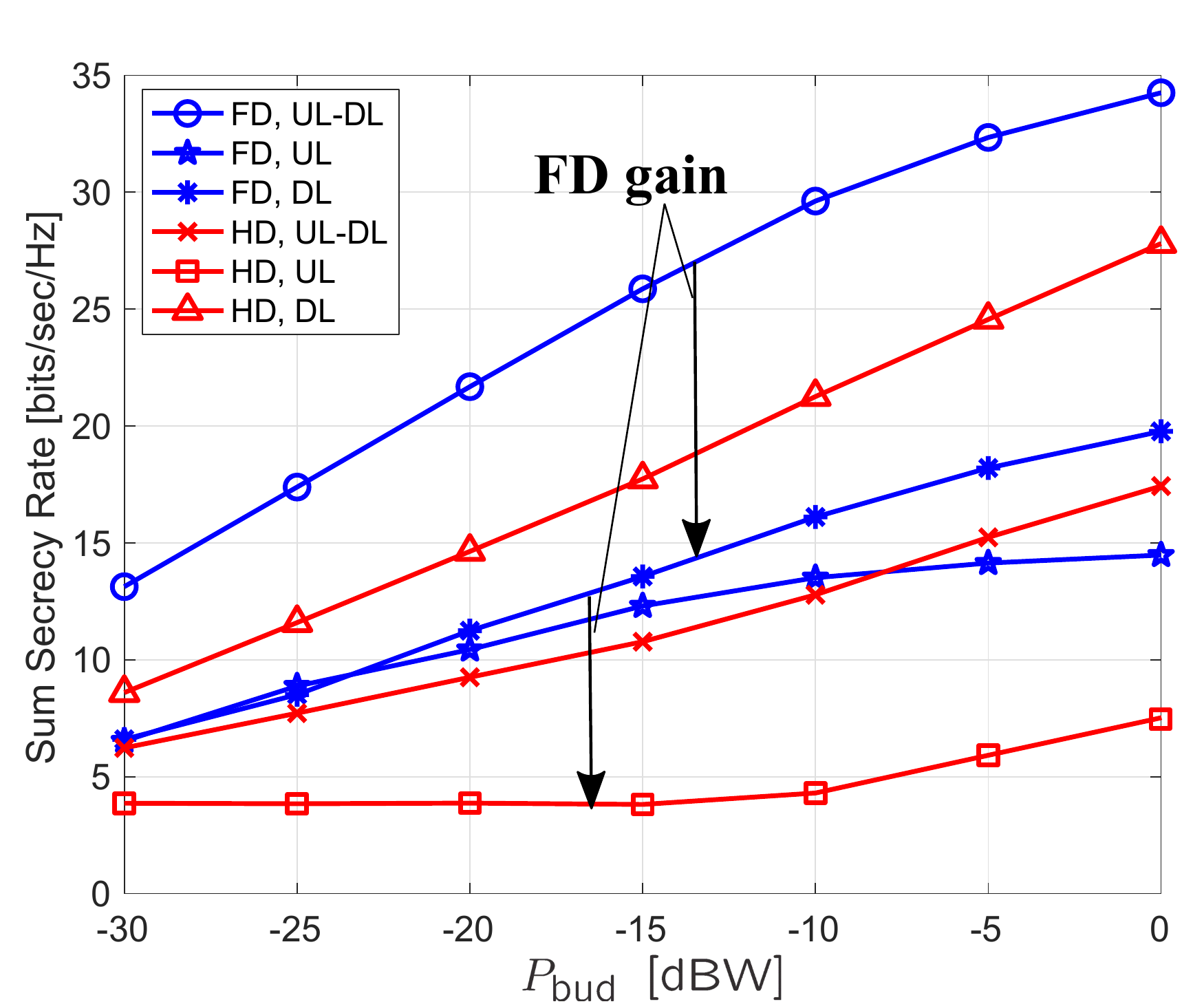}} 
\hspace{0cm}\subfigure[Sum Secrecy Rate vs. $P_{\text{bud}}$]{\includegraphics[height =  0.45\columnwidth, width = 0.5\columnwidth]{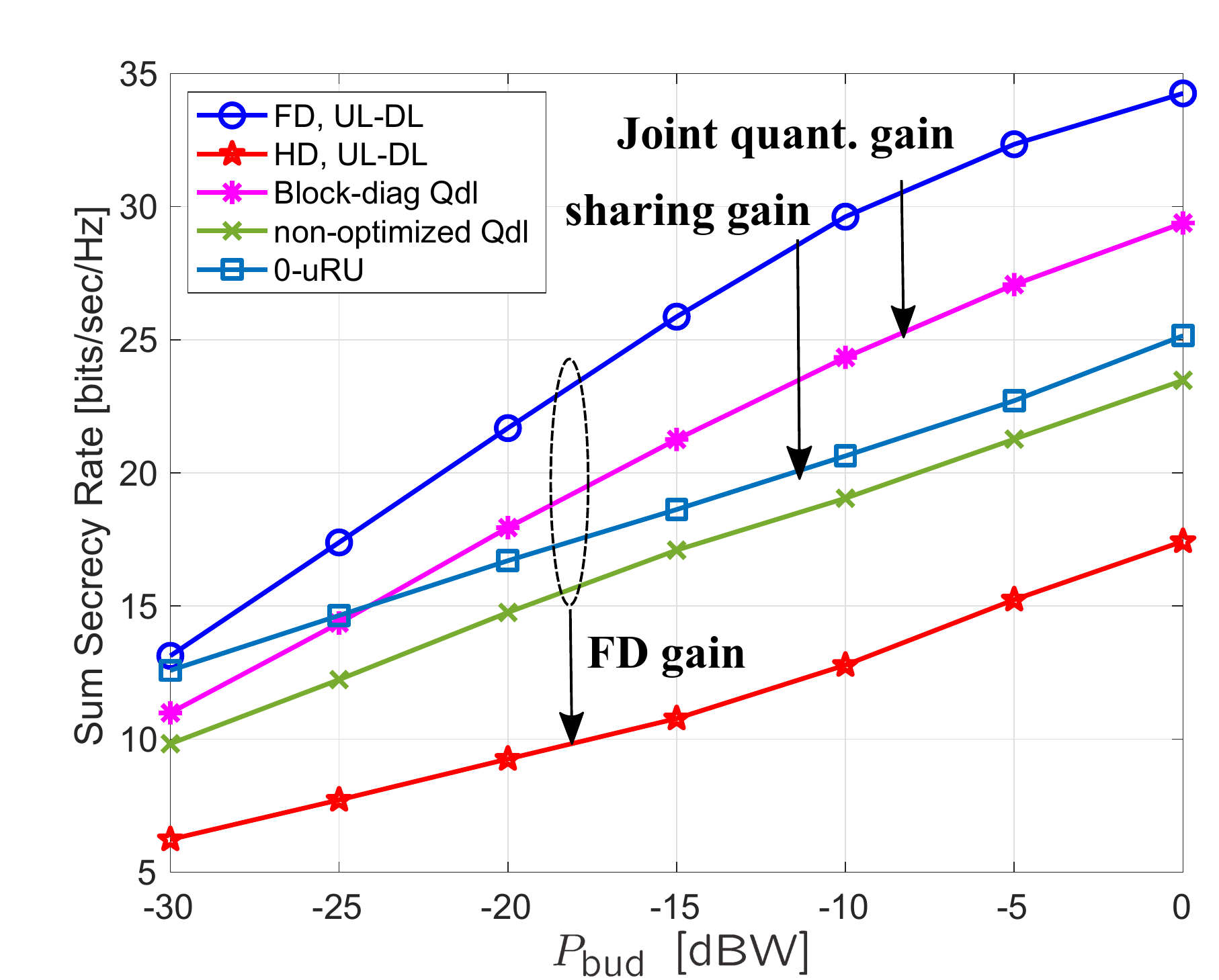}}   
    \caption{\textit{\textbf{(a)}}: Achieved secrecy spectral efficiency for UL and DL directions for different levels of Tx power budget. The gains of FD operation at the RUs is observed for both UL and sum UL-DL evaluations. \textit{\textbf{(b)}}: the gains of joint statistical DL quantization shaping as well as the sharing gain is observed via the proposed design.  } \label{fig_power}
\end{figure}

\begin{figure}[!t]  
\hspace{0cm}\subfigure[Sum Secrecy Rate vs. thermal noise variance]{\includegraphics[height =  0.45\columnwidth, width = 0.5\columnwidth]{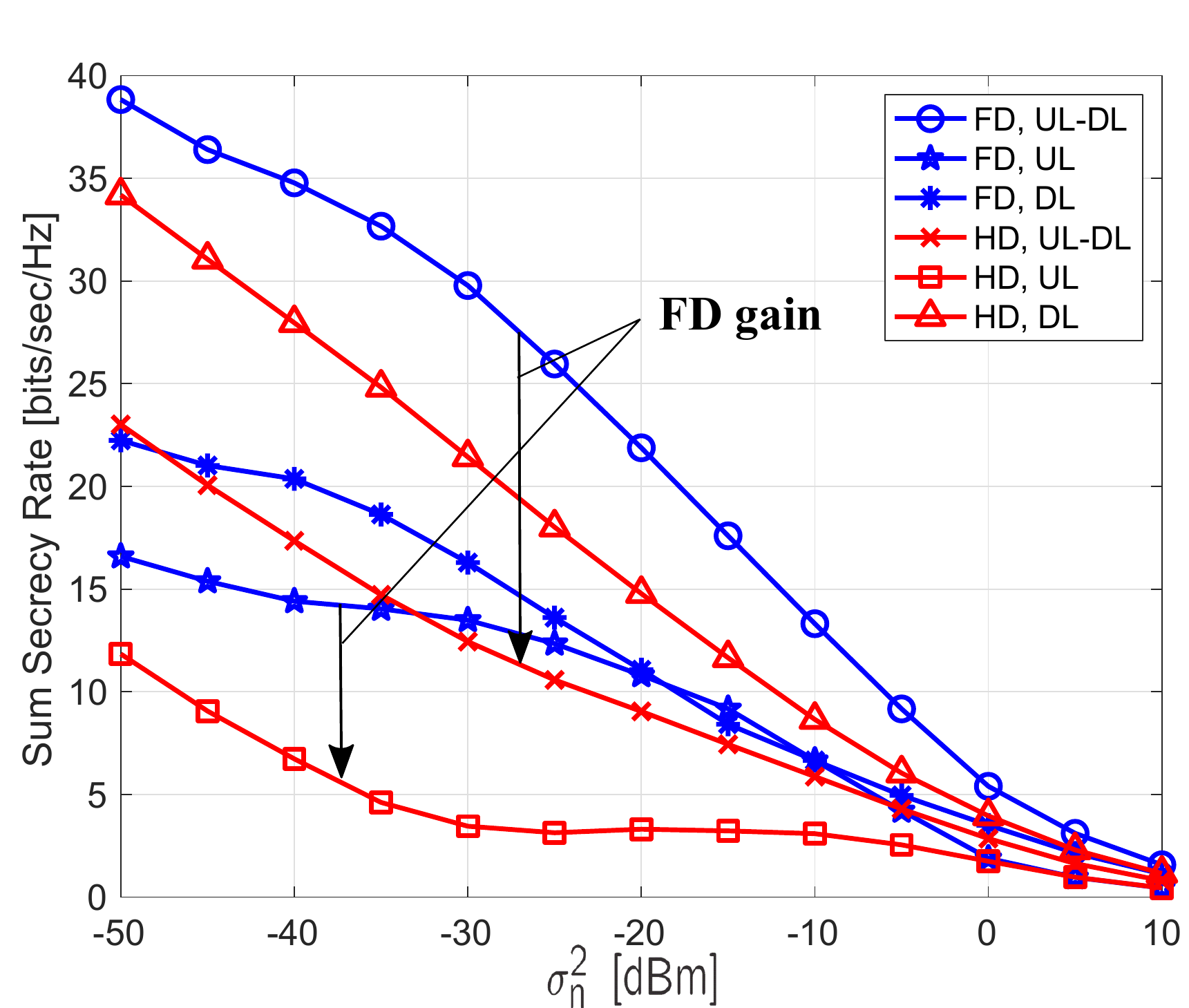}} 
\hspace{0cm}\subfigure[Sum Secrecy Rate vs. thermal noise variance]{\includegraphics[height =  0.45\columnwidth, width = 0.5\columnwidth]{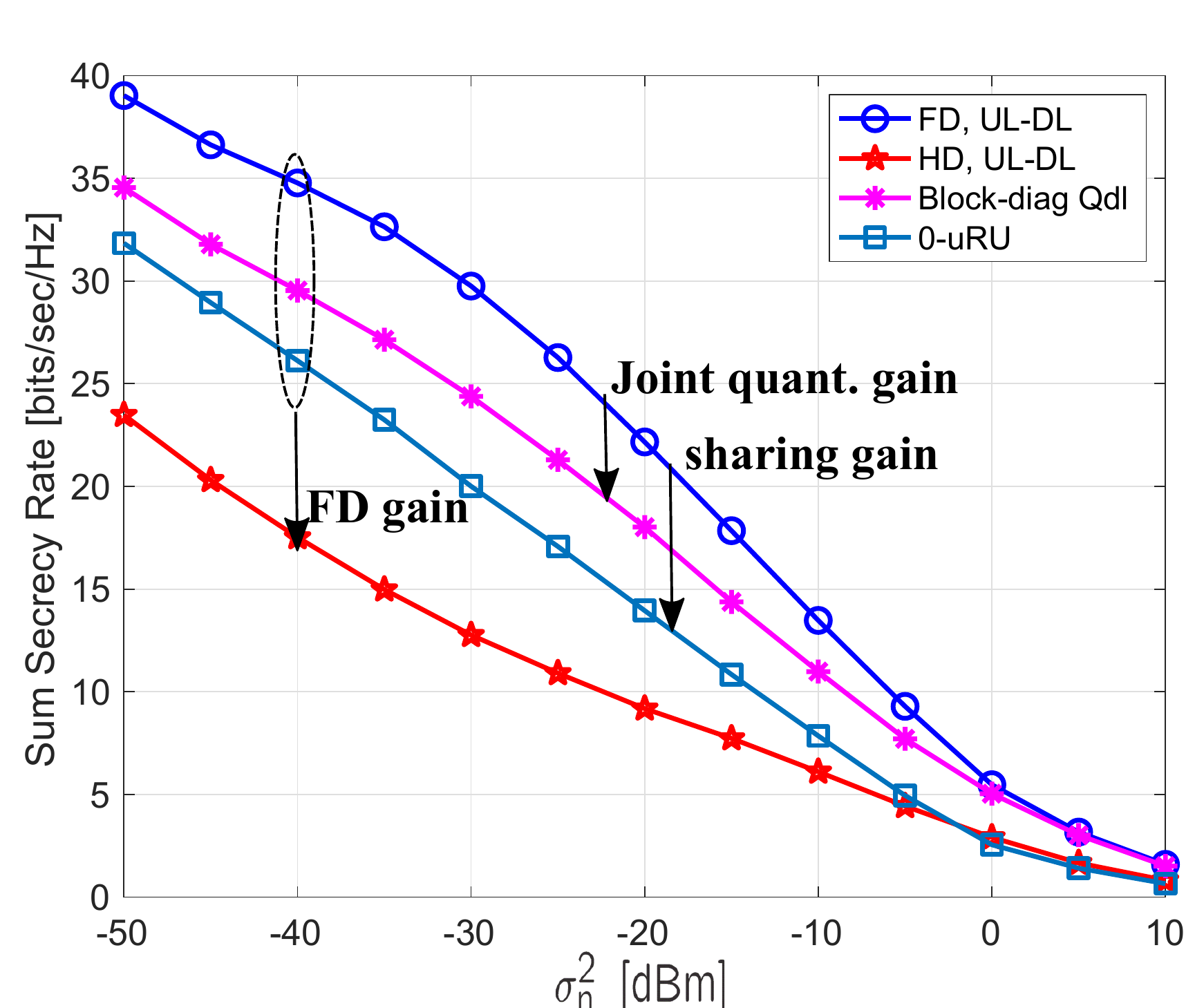} }   
    \caption{\textit{\textbf{(a)}}: Achieved secrecy spectral efficiency for UL and DL directions for different levels of thermal noise variance. The gains of FD operation at the RUs is observed for both UL and sum UL-DL for different noise levels. \textit{\textbf{(b)}}: the gains of joint statistical DL quantization shaping as well as the sharing gain is observed via the proposed design. } \label{fig_noise}
\end{figure}

\begin{figure*}[!h]  
\hspace{0cm}\subfigure[Sum Secrecy Rate vs. $\kappa$]{\includegraphics[height =  0.45\columnwidth, width = 0.5\columnwidth]{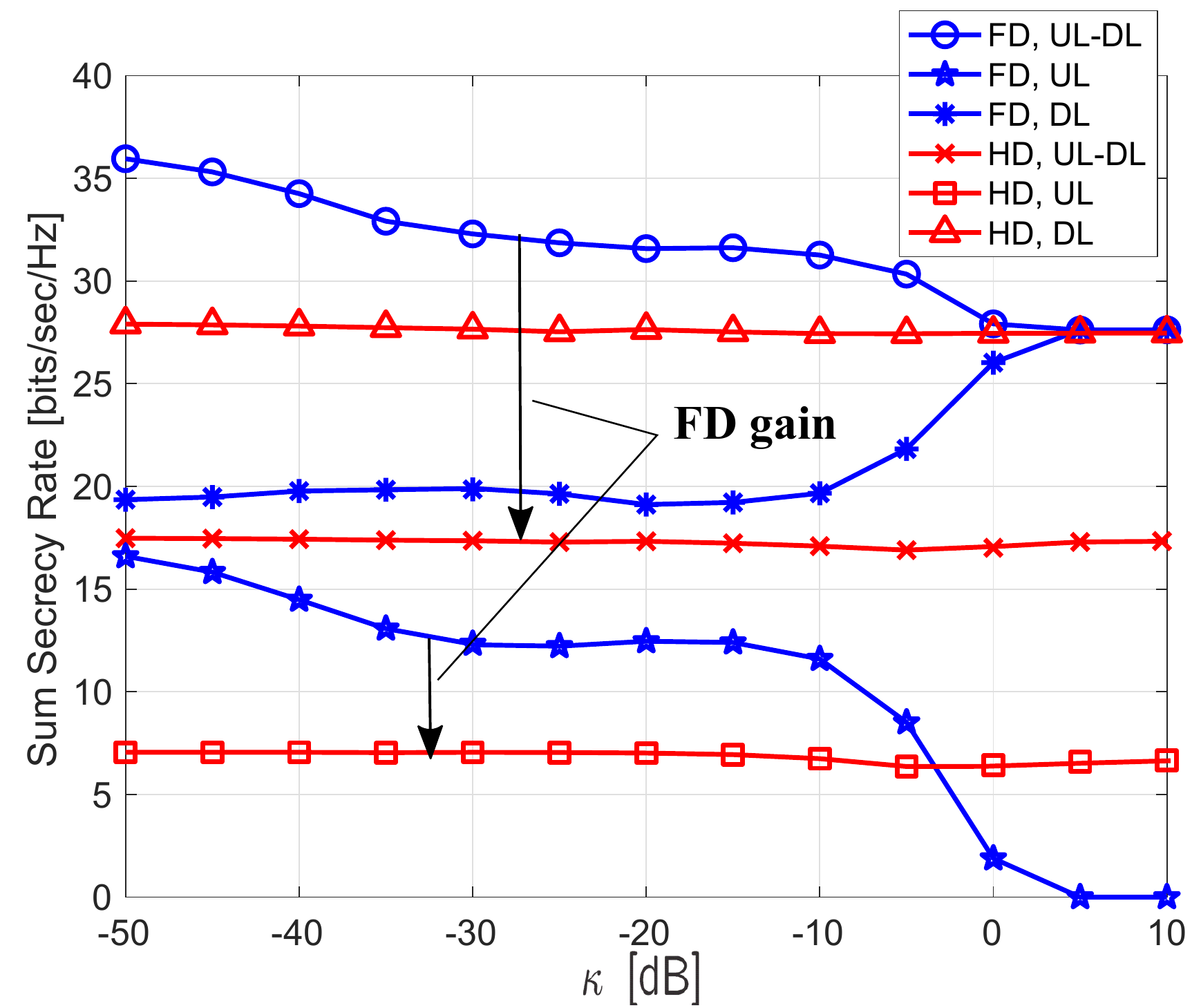} } 
\hspace{0cm}\subfigure[Sum Secrecy Rate vs. $\kappa$]{\includegraphics[height =  0.45\columnwidth, width = 0.5\columnwidth]{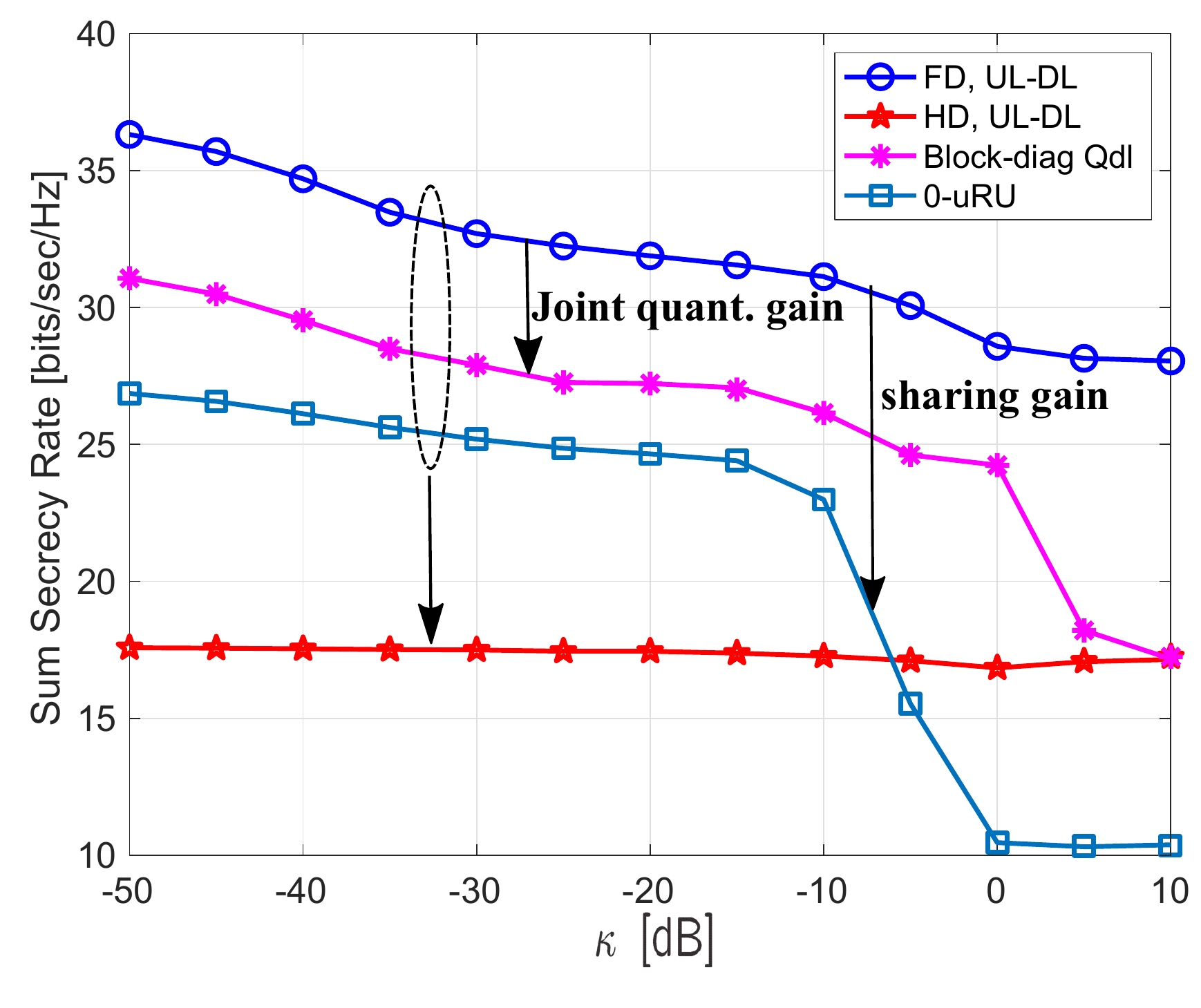}}   
    \caption{\textit{\textbf{(a)}}: Achieved secrecy spectral efficiency for UL and DL directions for different levels of transceiver accuracy. The gains of FD operation at the RUs are observed for both UL and sum UL-DL for different $\kappa$ levels. \textit{\textbf{(b)}}: the gains of joint statistical DL quantization shaping as well as the sharing gain is observed via the proposed design.} \label{fig_kappa}
\end{figure*} 

\section{Simulation Results}
In this section, we evaluate the performance of the studied system via numerical simulations. Please note that the proposed scheme enables secure sharing of the communications infrastructure, i.e., RU nodes, employing the FD capability of the RUs as well as the proposed statistical quantization shaping mechanism. In this regard, this is of interest to evaluate, firstly, the achievable gains as a result of the FD operation at the RU nodes, i.e., the secrecy spectral efficiency \textit{gain obtained via the coexistence of the UL and DL communications at the same channel} as well as the \textit{joint utilization of the fronthaul quantization for improving secrecy}, and second, the \textit{gain obtained via the secure sharing mechanism}, i.e., enabling the untrusted RUs to participate in the communication process without the loss of information privacy. 

We assume that the UL and DL users are uniformly distributed in a squared area of $100$ meters in length, where $4$ RUs are positioned each at the center of $4$ equally divided squares each with the length of $50$ meters. Among the deployed RUs, it is assumed that $2$ belong to a friendly operator, i.e., trusted RUs, and $2$ belong to an external operator or private owners, i.e., untrusted RUs. The trusted and untrusted RUs are positioned at opposite diagonals of the square cell, see Fig.~\ref{fig_general}-(a). Similarly as in \cite{8437192}, the channel between two different nodes with the distance $d$ is modeled as $\mathbf{H} = \sqrt{\rho}\tilde{\mathbf{H}}$, where $\rho=1/(1+(d/50)^3)$ represents the path-loss and $\text{vec}\left(\tilde{\mathbf{H}}\right)\sim\mathcal{CN}(\mathbf{0}, \mathbf{I})$. The self-interference channels are modeled similar to \cite{6353396} as 
\begin{align}
\mathbf{H}_{ii} \sim \mathcal{CN}\left(\sqrt{\frac{\rho_{\text{si}}K_{R}}{1+K_R}}\mathbf{H}_0,\frac{\rho_{\text{si}}}{1+K_R}\mathbf{I}_{M_{\text{R},i}}\otimes\mathbf{I}_{N_{\text{R},i}}\right), \forall i\in\mathcal{R}, \nonumber
\end{align}
where $\rho_{\text{si}}$ is the self-interference channel strength, $\mathbf{H}_0$ is a deterministic term indicating the dominant interference path\footnote{For simplicity, we choose $\mathbf{H}_0$ as a matrix of all-$1$ elements}, and $K_R = 10$ is the Rician coefficient. The resulting system performance corresponding to each parameter value and a specific implementation is then averaged over {$200$ channel realizations.}
 Unless otherwise is stated, the following are set as the default system parameters: $\left| \mathcal{R} \right|=4$, $\left| \mathcal{M} \right|=2$,  $\left| \mathcal{K} \right|=2$, $\rho_{\text{si}} = 1$, $N_{\text{U},k} = 2$, $N_{\text{R},m}= M_{\text{R},m} = 2$, $C_m = 100$ Mbit/s, $B = 10$ MHz, $P_{\text{bud}} = P_{\text{U},k} = P_{\text{R},m} = 30\;\text{[dBm]}$, $w_m = \bar{w}_k = 1$,  $\sigma_{\text{n}}^2 = N_{\text{ul},k}= N_{\text{dl},m} = -40\;\text{[dBm]}$, $\kappa=\beta = -40\; \text{[dB]}$, $\forall k \in \mathcal{K}, m \in \mathcal{R}$.  \par

In Fig.~\ref{fig_general}-(a), the simulated network setup is depicted. The solid black (gray) squares represent the deployed trusted (untrusted) RU nodes. The users and the potential eavesdropper nodes are distributed randomly within the cell area where the solid black and solid gray circles respectively represent the users and the potential eavesdroppers. As previously mentioned, the users and the RUs are distributed within a square cell area of $100$ meters length. 

In Fig.~\ref{fig_general}-(b), the average convergence behavior of Algorithm~1, as well as the proposed rank-reduction method is depicted for different values of transmit power level. Note that due to its iterative nature, the convergence behavior of Algorithm~1 is important as a measure of the required computational efforts, as well as to verify the expected monotonic improvement. It is observed that the algorithm converges within $100$ iterations. Moreover, it is observed that the proposed sequential rank-reduction method converges to a close proximity of the celebrated randomization method without the need to perform the costly re-adjustments, i.e., to re-run a reduced form of the optimization problem for a large number of randomization efforts in order to re-adjust the resulting instances from GR into the feasible solution space. Please note that while the algorithm convergence is reached in $40-70$ number of iterations, the algorithm reaches a close proximity of the eventual performance within $10-20$ iterations, which may also serve as a sub-optimal solution but with less computational cost.

In Fig.~\ref{fig_power} the secrecy performance of the proposed scheme is evaluated for different levels of transmit power as well as different implementation strategies. In particular, Fig.~\ref{fig_power}-(a) evaluates the secrecy rate performance in the DL, UL directions, as well as the sum secrecy rate performance, when RU nodes operate in FD and HD modes. The labels \textit{\textbf{``HD, UL"}}, \textit{\textbf{``HD, DL"}}, respectively represent the achieved secrecy spectral efficiency of an equivalent HD network in the UL and DL directions, whereas the label \textit{\textbf{``HD, UL-DL"}} represents the obtained sum spectral efficiency when TDD is utilized to accommodate UL and DL link directions in different channel resources. Similarly, the labels \textit{\textbf{``FD, UL"}}, \textit{\textbf{``FD, DL"}}, and \textit{\textbf{``FD, UL-DL"}} represent the obtained spectral efficiency in a network with FD capability associated with the UL, DL and all link directions. It is observed that the proposed quantization-aided FD jamming leads to both a higher sum secrecy spectral efficiency, as well as a significantly higher secrecy rate in the UL direction. This is expected, since for an FD RU, the DL fronthaul quantization simultaneously acts as the jamming signal on the untrusted RUs in the DL direction, as well as the RUs for the UL transmission from the users. Nevertheless, while the DL fronthaul quantization is utilized also for an HD network for the purpose of DL information secrecy, it provides no mechanism for protecting the UL information against the untrusted RUs. 

In Fig.~\ref{fig_power}-(b), in addition to the observed gain by employing FD operation at the RUs in Fig.~\ref{fig_power}-(a), the significance of the proposed joint quantization covariance shaping is evaluated. The benchmarks with the label \textit{\textbf{``non-optimized Qdl"}} and \textit{\textbf{``Block-diag Qdl"}}, respectively represent the scenarios where the DL quantization is not optimized for the purpose of secrecy enhancement, i.e., is not directed/shaped for protecting UL/DL information from the untrusted entities, and the scenario where the quantization statistics is not jointly shaped at all RUs, i.e., the DL quantization covariance is shaped separately at each RU, which results in a block-diagonal $\ma{Q}_{\text{dl}}$. It is observed that the implemented schemes enjoy a notable gain by jointly shaping and optimizing the DL quantization noise at all RUs, which acts as a key mechanism for information protection in both UL and DL directions. In addition to the impact of the optimized DL quantization shaping, the benchmark with the label \textit{\textbf{``0-uRU"}} evaluates the scenario where the untrusted RUs are merely treated as traditional eavesdroppers and not used for the purpose of UL/DL communication. Please note that the latter case represents the traditional scenario, where the untrusted entities are merely ignored or treated as eavesdroppers, but not constructively used in the communication process. In this respect, the proposed information secrecy mechanism offers a \textit{sharing mechanism} where the RU infrastructure nodes belonging to a private owner or exotic operators can be integrated as part of the desired communication process, while preserving the information privacy requirements.       

In Fig.~\ref{fig_noise} the secrecy performance of the proposed scheme is evaluated for different levels of thermal noise. As expected, it is observed that a higher level of thermal noise variance degrades the secrecy spectral efficiency in all directions and for different implementation strategies. In particular, it is observed from Fig.~\ref{fig_noise}-(a) that the FD secrecy gain due to the UL and DL coexistence is preserved also for the high thermal noise regimes, wheres the gains obtained by the quantization shaping mechanism is degraded as the thermal noise increases, see Fig.~\ref{fig_noise}-(b). This is expected, as the high thermal noise level leads to a reduction in the significance of the self-interference and the co-channel interference, which are the degrading factors for an FD system performance compared to an HD one. However, as the variance of the thermal noise increases, the thermal noise leads to a natural jamming effect on the undesired receivers, as it degrades the decoding capability at each individual RU. Nevertheless, it is observed from Fig.~\ref{fig_noise}-(b) that the associated gains with the joint quantization shaping and FD operations are especially significant in higher signal-to-noise regimes, which emphasizes the significance of the proposed scheme in the favorable scenarios. 

In Fig.~\ref{fig_kappa} the secrecy performance is depicted for different levels of the self-interference cancellation quality. Please note that the proposed scheme heavily relies on the FD operation at the trusted entities to enable the co-utilization of the quantization noise also as a jamming signal to protect information in the UL and DL directions. Nevertheless, the FD operation, after the utilization of the state-of-the-art self-interference cancellation methods, leads to an increase in the receiver impairments due to the residual self-interference. In this regard, it is observed from Fig.~\ref{fig_kappa} that the promising gain of the proposed scheme in the secrecy performance vanishes and converges to the performance of an equivalent HD system for the large impairment levels, i.e., high $\kappa$. This is expected, as a large value of $\kappa$ (or, equivalently, a poor self-interference cancellation quality) forces the system to operate in the HD mode in order to avoid large residual self-interference. Interestingly, while the behavior of the HD system remains almost constant in the face of different levels of $\kappa$, it is observed that the DL performance of an FD system improves as $\kappa$ increases, whereas the UL performance reaches close to zero. This is since, for an FD system with poor self-interference cancellation quality, the UL communications face with the strong residual self-interference. Hence, the communication in the UL direction is usually turned off at a high $\kappa$ regime in order to reduce the co-channel interference effect in the DL direction.

\section{Conclusion}
In this work, we have proposed a mechanism for ensuring information secrecy in both UL and DL directions in an FD C-RAN, utilizing the DL quantization noise also as a jamming signal towards different untrusted entities. The key take-aways of this work can be summarized as follows. \textit{\textbf{Firstly}}, for a traditional system without a jamming or quantization-aided secrecy mechanism, it is observed that guaranteeing information privacy in the physical layer leads to a severe performance loss and resource inefficiency, considering the large margin of performance degradation when the system is not adjusted for secrecy improvement. \textit{\textbf{Secondly}}, a significant gain is observed via the application of the proposed secrecy-enhancing mechanism, however, the \textit{\textbf{secrecy-aware quantization gain}} is highly influenced by the accuracy of the FD transceivers due to the degrading impact of residual self-interference. \textit{\textbf{Thirdly}}, a promising gain can be obtained in the achievable sum secrecy rate via the participation of the external/untrusted RUs, i.e., \textit{\textbf{sharing gain}}, when the proposed quantization-aided jamming strategy is implemented in a system with a high transceiver dynamic range.

\end{document}